\documentclass[apjl]{emulateapj}
\usepackage{comment}
\usepackage{ifthen}

\newcommand{\ctbd}[1]{}

\newcommand{\lc}{light curve}
\newcommand{\lcs}{light curves}
\newcommand{\Lc}{Light curve}

\newcommand{\ccdsize}[1]{\ensuremath{\rm #1\times\rm#1}}
\newcommand{\fovsize}[2]{\ensuremath{\rm #1 #2\times\rm#1 #2}}

\newcommand{\band}[1]{\ensuremath{#1}-band}
\newcommand{\ordo}{\ensuremath{\mathcal{O}}}
\newcommand{\chisq}{\ensuremath{\chi^2}}

\newcommand{\kms}{\ensuremath{\rm km\,s^{-1}}}
\newcommand{\ms}{\ensuremath{\rm m\,s^{-1}}}

\newcommand{\gcmc}{\ensuremath{\rm g\,cm^{-3}}}
\newcommand{\ergscmsq}{\ensuremath{\rm erg\,s^{-1}\,cm^{-2}}}

\newcommand{\pxs}{\ensuremath{\rm \arcsec pixel^{-1}}}

\newcommand{\masyr}{\ensuremath{\rm mas\,yr^{-1}}}

\newcommand{\mplsini}{\ensuremath{\mpl\sin i}}

\newcommand{\vsini}{\ensuremath{v \sin{i}}}
\newcommand{\feh}{\ensuremath{\rm [Fe/H]}}

\newcommand{\Savg}{\ensuremath{\langle S\rangle}}

\newcommand{\rsun}{\ensuremath{R_\sun}}
\newcommand{\msun}{\ensuremath{M_\sun}}
\newcommand{\lsun}{\ensuremath{L_\sun}}

\newcommand{\rstar}{\ensuremath{R_\star}}
\newcommand{\mstar}{\ensuremath{M_\star}}
\newcommand{\lstar}{\ensuremath{L_\star}}

\newcommand{\teffstar}{\ensuremath{T_{\rm eff\star}}}
\newcommand{\rhostar}{\ensuremath{\rho_\star}}
\newcommand{\loggstar}{\ensuremath{\log{g_{\star}}}}

\newcommand{\rearth}{\ensuremath{R_\earth}}
\newcommand{\mearth}{\ensuremath{M_\earth}}

\newcommand{\rpl}{\ensuremath{R_{p}}}
\newcommand{\mpl}{\ensuremath{M_{p}}}

\newcommand{\rhopl}{\ensuremath{\rho_{p}}}

\newcommand{\loggpl}{\ensuremath{\log g_{p}}}

\newcommand{\arstar}{\ensuremath{a/\rstar}}
\newcommand{\zrstar}{\ensuremath{\zeta/\rstar}}

\newcommand{\rjup}{\ensuremath{R_{\rm J}}}
\newcommand{\mjup}{\ensuremath{M_{\rm J}}}

\newcommand{\reffig}[1]{Fig.~\ref{fig:#1}}
\newcommand{\refsec}[1]{\mbox{\S\ \ref{sec:#1}}}
\newcommand{\refeq}[1]{Eq.~\ref{eq:#1}}
\newcommand{\reftab}[1]{Tab.~\ref{tab:#1}}

\newcommand{\flwof}{\mbox{FLWO 1.2\,m}}

\newcommand{\ssts}{{\em Spitzer}}

\newcommand{\hst}{{\em HST}}

\newcommand{\piszkessch}{Konkoly 0.6\,m Schmidt}

\newcommand{\dscu}{\mbox{$\delta$ Scuti}}

\newcommand{\hd}[1]{\mbox{HD #1}}

\newcommand{\hip}[1]{\mbox{HIP #1}}

\newcommand{\gj}[1]{\mbox{GJ #1}}

\newboolean{emulateapj}
\setboolean{emulateapj}{true}

\newboolean{astroph}
\setboolean{astroph}{true}

\ifthenelse{\boolean{emulateapj}}{
	\newcommand{\titledag}{$\dagger$}
}{
	\newcommand{\titledag}{\dagger}
}

\newcommand{\hatcur}{HAT-P-11}
\newcommand{\hatcurb}{HAT-P-11b}

\newcommand{\hatcurCCra}{19h50m50.21s}                                 
\newcommand{\hatcurCCdec}{+48d04m50.8s}                                
\renewcommand{\hatcurCCra}{ \ensuremath{ 19^{\mathrm h}50^{\mathrm m}50.21^{\mathrm s}}}  
\renewcommand{\hatcurCCdec}{\ensuremath{+48^{\mathrm d}04^{\mathrm m}50.8^{\mathrm s}} }  
\newcommand{\hatcurCCmag}{9.59}                                       
\newcommand{\hatcurCCtwomass}{2MASS~19505021+4804508}                  
\newcommand{\hatcurCCgsc}{GSC~03561-02092}                             
\newcommand{\hatcurCCtassmv}{9.587}                                    
\newcommand{\hatcurLCdip}{\ensuremath{4.2}}                            
\newcommand{\hatcurLCrprstar}{\ensuremath{0.0576\pm0.0009}}            
\newcommand{\hatcurLCimp}{\ensuremath{0.347_{-0.139}^{+0.130}}}        
\newcommand{\hatcurLCdur}{\ensuremath{0.0957\pm0.0012}}                
\newcommand{\hatcurLCingdur}{\ensuremath{0.0051\pm0.0013}}             
\newcommand{\hatcurLCP}{\ensuremath{4.8878162\pm0.0000071}}            
\newcommand{\hatcurLCPprec}{\ensuremath{4.8878162}}                    
\newcommand{\hatcurLCPshort}{\ensuremath{4.8878}}                      
\newcommand{\hatcurLCT}{\ensuremath{2454605.89132\pm0.00032}}          
\newcommand{\hatcurLCzeta}{\ensuremath{22.15\pm0.15}}                  
\newcommand{\hatcurLCbb}{\ensuremath{0.120\pm0.087}}                   
\newcommand{\hatcurSMEteff}{\ensuremath{4780\pm50}}                    
\newcommand{\hatcurSMEzfeh}{\ensuremath{+0.31\pm0.05}}                  
\newcommand{\hatcurSMEvsin}{\ensuremath{1.5\pm1.5}}                    
\newcommand{\hatcurYYm}{\ensuremath{0.81_{-0.03}^{+0.02}}}             
\newcommand{\hatcurYYmlong}{\ensuremath{0.809_{-0.027}^{+0.020}}}      
\newcommand{\hatcurYYr}{\ensuremath{0.75\pm0.02}}                      
\newcommand{\hatcurYYrlong}{\ensuremath{0.752\pm0.021}}                
\newcommand{\hatcurYYlogg}{\ensuremath{4.59\pm0.03}}                   
\newcommand{\hatcurYYlum}{\ensuremath{0.26\pm0.02}}                    
\newcommand{\hatcurYYmv}{\ensuremath{6.57\pm0.09}}                     
\newcommand{\hatcurYYage}{\ensuremath{6.5_{-4.1}^{+5.9}}}              
\newcommand{\hatcurYYspec}{K4}                                         
\newcommand{\hatcurRVK}{\ensuremath{11.6\pm1.2}}                       
\newcommand{\hatcurRVk}{\ensuremath{0.201\pm0.049}}                    
\newcommand{\hatcurRVh}{\ensuremath{0.051\pm0.092}}                    
\newcommand{\hatcurRVgamma}{\ensuremath{-0.4\pm0.8}}                   
\newcommand{\hatcurPPi}{\ensuremath{88.5\pm0.6}}                       
\newcommand{\hatcurPPlogg}{\ensuremath{3.05\pm0.06}}                   
\newcommand{\hatcurPPar}{\ensuremath{15.58_{-0.82}^{+0.17}}}           
\newcommand{\hatcurPParel}{\ensuremath{0.0530_{-0.0008}^{+0.0002}}}    
\newcommand{\hatcurPPrho}{\ensuremath{1.33\pm0.20}}                  
\newcommand{\hatcurPPmlong}{\ensuremath{0.081\pm0.009}}                
\newcommand{\hatcurPPrlong}{\ensuremath{0.422\pm0.014}}                
\newcommand{\hatcurPPmrcorr}{\ensuremath{0.025}}                       
\newcommand{\hatcurPPteff}{\ensuremath{878\pm15}}                      
\newcommand{\hatcurPPtheta}{\ensuremath{0.025\pm0.003}}                
\newcommand{\hatcurRVkcorr}{\ensuremath{0.190\pm0.046}}                
\newcommand{\hatcurRVhcorr}{\ensuremath{-0.016\pm0.056}}               
\newcommand{\hatcurRVckh}{\ensuremath{0.931\pm0.063}}                  
\newcommand{\hatcurRVeccen}{\ensuremath{0.198\pm0.046}}                
\newcommand{\hatcurRVomega}{\ensuremath{355.2\pm17.3}}                  
\newcommand{\hatcurXsecondary}{\ensuremath{2454608.96\pm0.15}}         
\newcommand{\hatcurXsecdur}{\ensuremath{0.1006\pm0.0130}}              
\newcommand{\hatcurXsecingdur}{\ensuremath{0.0054\pm0.0013}}           
\newcommand{\hatcurPPfluxperi}{\ensuremath{2.04\cdot10^{8}\pm2.78\cdot10^{7}}}       
\newcommand{\hatcurPPfluxap}{\ensuremath{9.11\cdot10^{7}\pm9.53\cdot10^{6}}}         
\newcommand{\hatcurPPfluxavg}{\ensuremath{1.34\cdot10^{8}\pm9.38\cdot10^{6}}}        
\newcommand{\hatcurXdist}{\ensuremath{38.0\pm1.3}}                     

\newcommand{\hatcurCCpm}{\ensuremath{263.32\pm1.31}}                   

\newcommand{\hatcurPPme}{\ensuremath{25.8\pm2.9}}                      
\newcommand{\hatcurPPmelong}{\ensuremath{25.8\pm2.9}}                  
\newcommand{\hatcurPPre}{\ensuremath{4.73\pm0.16}}                     
\newcommand{\hatcurPPrelong}{\ensuremath{4.73\pm0.16}}                 

\newcommand{\hatcurRVlindrift}{\ensuremath{0.0297 \pm 0.0050}}         
\newcommand{\hatcurSMEzfehNE}{\ensuremath{+0.31}}                       

\shortauthors{Bakos et al.}
\shorttitle{HAT-P-11b}

\begin{document}

\title{\hatcur\lowercase{b}: A Super-Neptune Planet Transiting a Bright K
Star in the Kepler Field\altaffilmark{\titledag}}

\author{
	G.~\'A.~Bakos\altaffilmark{1,2},
	G.~Torres\altaffilmark{1},
	A.~P\'al\altaffilmark{1,4},
	J.~Hartman\altaffilmark{1},
	G\'eza~Kov\'acs\altaffilmark{3},
	R.~W.~Noyes\altaffilmark{1},
	D.~W.~Latham\altaffilmark{1},
	D.~D.~Sasselov\altaffilmark{1},
	B.~Sip\H{o}cz\altaffilmark{1,4},
	G.~A.~Esquerdo\altaffilmark{1},
	D.~A.~Fischer\altaffilmark{5},
	J.~A.~Johnson\altaffilmark{6},
	G.~W.~Marcy\altaffilmark{7},
	R.~P.~Butler\altaffilmark{8},
	H.~Isaacson\altaffilmark{5},
	A.~Howard\altaffilmark{7},
	S.~Vogt\altaffilmark{9},
	G\'abor~Kov\'acs\altaffilmark{1},
	J.~Fernandez\altaffilmark{1},
	A.~Mo\'or\altaffilmark{3},
	R.~P.~Stefanik\altaffilmark{1},
	J.~L\'az\'ar\altaffilmark{9},
	I.~Papp\altaffilmark{9},
	P.~S\'ari\altaffilmark{9}
}

\altaffiltext{1}{Harvard-Smithsonian Center for Astrophysics,
	Cambridge, MA, gbakos@cfa.harvard.edu}

\altaffiltext{2}{NSF Fellow}

\altaffiltext{3}{Konkoly Observatory, Budapest, Hungary}

\altaffiltext{4}{Department of Astronomy,
	E\"otv\"os Lor\'and University, Budapest, Hungary.}

\altaffiltext{5}{Department of Physics and Astronomy, San Francisco
	State University, San Francisco, CA}

\altaffiltext{6}{Institute for Astronomy, University of Hawaii,
	Honolulu, HI 96822; NSF Postdoctoral Fellow}

\altaffiltext{7}{Department of Astronomy, University of California,
	Berkeley, CA}

\altaffiltext{8}{Department of Terrestrial Magnetism, Carnegie
	Institute of Washington, DC}

\altaffiltext{9}{University of California Observatories/Lick
	Observatory, University of California at Santa Cruz, Santa Cruz, CA
	95064}

\altaffiltext{10}{Hungarian Astronomical Association, Budapest, 
	Hungary}

\altaffiltext{$\dagger$}{%
	Based in part on observations obtained at the W.~M.~Keck
	Observatory, which is operated by the University of California and
	the California Institute of Technology. Keck time has been
	granted by NOAO (A285Hr) and NASA (N128Hr).
}


\begin{abstract}

We report on the discovery of \hatcurb{}, the smallest radius
transiting extrasolar planet (TEP) discovered from the ground, 
and the first hot Neptune discovered to date by transit searches.
\hatcurb\ orbits the bright (V=\hatcurCCtassmv) and metal rich ($\feh =
\hatcurSMEzfeh$) \hatcurYYspec\ dwarf star \hatcurCCgsc\ with $P =
\hatcurLCP$~days and produces a transit signal with depth of
\hatcurLCdip\,mmag; the shallowest found by transit searches that is
due to a confirmed planet. We present a global analysis of the
available photometric and radial-velocity data that result in stellar
and planetary parameters, with simultaneous treatment of systematic
variations.  The planet, like its near-twin \gj{436b}, is somewhat
larger than Neptune (17\mearth, 3.8\rearth) both in mass $\mpl =
\hatcurPPmlong\,\mjup\ (\hatcurPPme\,\mearth)$ and radius $\rpl =
\hatcurPPrlong\,\rjup\ (\hatcurPPre\,\rearth)$. \hatcurb\ orbits in an
eccentric orbit with $e = \hatcurRVeccen$ and $\omega =
\hatcurRVomega\arcdeg$, causing a reflex motion of its parent star with
amplitude \hatcurRVK\,\ms, a challenging detection due to the high
level of chromospheric activity of the parent star.  Our ephemeris for
the transit events is $T_c = \hatcurLCT$ (BJD), with duration
\hatcurLCdur\,d, and secondary eclipse epoch of \hatcurXsecondary\,d
(BJD). The basic stellar parameters of the host star are $\mstar =
\hatcurYYmlong\,\msun$, $\rstar = \hatcurYYrlong\,\rsun$ and $\teffstar
= \hatcurSMEteff\,K$.  Importantly, \hatcur\ will lie on one of the
detectors of the forthcoming Kepler mission; this should make possible
fruitful investigations of the detailed physical characteristic of both
the planet and its parent star at unprecedented precision. We discuss
an interesting constraint on the eccentricity of the system by the
transit \lc\ and stellar parameters. This will be particularly useful
for eccentric TEPs with low amplitude RV variations in Kepler's field.
We also present a blend analysis, that for the first time treats the
case of a blended transiting hot Jupiter mimicing a transiting hot
Neptune, and proves that \hatcurb\ is not such a blend.

\end{abstract}

\keywords{ 
	planetary systems ---
	stars: individual (\hatcur{}, \hatcurCCgsc{}) 
	techniques: spectroscopic, photometric
}


\section{Introduction}
\label{sec:intro}

Transiting extrasolar planets (TEPs) are uniquely valuable for
understanding the nature of planetary bodies, because the transit light
curve, combined with precise radial velocity (RV) measurements of the
reflex motion of the parent star, yield unambiguous information on the
true mass and radius of the planet, assuming that the stellar mass and
radius are known. By inference, it is then possible to investigate the
internal structure of these planets, as has been done by several teams
trying to formulate and match theories to the observed bulk properties
of known TEPs \citep[e.g.][and references
therein]{baraffe:2008,fortney:2007,burrows:2007,seager:2007}. The
transits across the face of the star enable a plethora of scientific
follow-up opportunities, such as detection of the atmospheres of these
planets via transmission spectroscopy \citep{dc:2002}, measurement of
the stellar spin axis versus planetary orbit
\citep{winn:2005,johnson:2008a} via the Rossiter-McLaughlin effect
\citep{rossiter:1924,mclaughlin:1924}, or measurement of their
equilibrium temperature while they are occulted by their central stars
\citep{dc:2005}.

Photometric searches for TEPs have published some 50 such objects over
the past 8 years\footnote{http://www.exoplanet.eu/catalog-transit.php},
most of these with masses and radii in excess of that of Jupiter.
Previously the smallest mass TEP discovered by the transit search
method was HAT-P-1b with $M = 0.52\mjup$ and $R = 1.22\rjup$
\citep{bakos:2007a}, recently superseded by the discovery of
WASP-11/HAT-P-10b with $M = 0.46\mjup$ \citep{west:2008,bakos:2008}.
The smallest radius planet detected by ground-based transit searches
was HAT-P-3b \citep{torres:2007} with $R = 0.89\rjup$, and the smallest
radius planet from the space is Corot-7b \citep{leger:2009}.

In the meantime radial velocity surveys have been reaching down to
Neptune-mass planets, thanks to high precision and high signal-to-noise
spectrographs that deliver radial velocities at the \ms\ level over an
extended time, such as HARPS \citep{mayor:2003} on the ESO 3.6\,m
telescope, or HIRES on Keck \citep{vogt:1994}. It was a major advance
when \citet{santos:2004} discovered the $\mplsini = 14\,\mearth$ planet
around \mbox{$\mu$ Ara}, \citet{mcarthur:2004} detected a Neptune-mass
planet around \mbox{55 Cnc}, and \citet{butler:2004} found a $\sim
21\mearth$ mass planet around \gj{436}. These were followed by further
exo-Neptune discoveries, such as the three Neptune planetary system
around \hd{69830} found by \citet{lovis:2006}.

Recently the detection threshold of RV searches has reached even below
that of super-Earths ($M\lesssim 10\mearth$). \citet{rivera:2005} found
a $\sim 7.5\,\mearth$ super-Earth orbiting the nearby M dwarf \gj{876}.
\citet{udry:2007} found a 5\,\mearth\ and an 8\,\mearth\ mass planet in
a triple planetary system around \gj{581}. Finally, \citet{mayor:2008}
discovered a triple super-Earth system with 4.2, 6.9, and 9.2 earth
masses around \hd{40307}. Altogether, as of writing, some 20 objects
with minimum mass $\mplsini<0.1\,\mjup = 31.8\,\mearth$ have been
detected by the RV technique.

Radial velocity detections are routinely checked for transit events by
the discovery teams, or by the {\tt transitsearch.org} collaboration of
amateur/professional astronomers \citep{seagroves:2003}. A few
successful detections have been reported\footnote{Of course, the first
and most prominent one being \hd{209458b} \citep{dc:2000,henry:2000}.}.
One such case is \hd{189733b}, a 1.13\,\mjup\ planet around a K dwarf,
discovered by \citet{bouchy:2005} via the RV method, and confirmed to
transit by the same team via the Rossiter-McLaughlin effect, and then
via follow-up photometric observations.  Another example is the
21.2~day period eccentric planet \hd{17156b} found by
\citet{fischer:2007}, with transits detected by \citet{barbieri:2007}
through {\tt transitsearch.org}. A third example, \hd{149026b}, is a
transition object between Jupiter-mass and Neptune-mass planets, in the
sense that it is a hot Saturn with $\mpl = 0.36\,\mjup$ (or about 1.2
times the mass of Saturn), and $\rpl = 0.71\,\rjup$. The RV detection by
\citet{sato:2005} was followed by discovery of the 0.3\% deep transits
by the same team.

Among the $\sim20$ Neptune-mass objects found by RV searches, only one
is known to transit. This is \gj{436b}, whose transits were recovered
by \citet{gillon:2007}. \gj{436b} is thus an extremely valuable and
unique object, the only Neptune-mass planet other than our own Uranus
and Neptune, where the radius has been determined \citep[$R \approx
4.9\,\rearth$ or 4.2\,\rearth; ][respectively, hereafter B08 and
T08]{bean:2008a,torres:2008}, and its internal structure investigated
\citep[e.g.][]{baraffe:2008}. Based on these results, \gj{436b} is a
super-Neptune with $\sim 22\,\mearth$ total mass with extreme heavy
element enrichment, and only $\sim 10\%$ mass contribution by a H/He
envelope.

One of the wide-field surveys involved in the detection of TEP's is the
HATNet survey \citep{bakos:2002,bakos:2004}, which currently operates 6
small fully-automated wide-field telescopes. One station is the Fred
Lawrence Whipple Observatory (FLWO) of the Smithsonian Astrophysical
Observatory (SAO) on Mt.~Hopkins in Arizona with four telescopes
(HAT-5, HAT-6, HAT-7, HAT-10), and the other is the rooftop of the
Submillimeter Array Hangar (SMA) of SAO atop Mauna Kea, Hawaii. These
telescopes are modest 0.11m diameter f/1.8 focal ratio telephoto lenses
that are using front-illuminated CCDs at 5-min integration times.

Here we report on HATNet's discovery of \hatcurb, the second transiting
hot Neptune known. \hatcurb\ orbits the bright (V=\hatcurCCmag) K dwarf
star \hatcurCCgsc{} (hereafter called HAT-P-11) with a period of
\hatcurLCPshort\,days, and when it transits the star it causes a dip in
the stellar light curve of about \hatcurLCdip\,millimag (mmag). 
\hatcurb\ is certainly the smallest radius planet found by ground-based transit
searches, the only planet known with smaller radius being Corot-7b
\citep{leger:2009}. Importantly, the coordinates of the parent star place it
on one of the detectors of the forthcoming Kepler mission; this should
allow a broad range of useful follow-on observations to characterize
both the planetary system and the parent star. Because \hatcur\ is a
bright star, but still conveniently below the bright limit of the
Kepler mission, the extraordinary precision of repeated measurements
made over several years should lead to very accurate characterization
of the system. We note that \hatcurb\ would have not been detected by
HATNet if it orbited a significantly earlier star, such as the typical
F and G dwarf stars making up the bulk of the HATNet transit
candidates. 

The layout of the paper is as follows. First we describe in
\refsec{obs} the observational data that led to the discovery of
\hatcurb, including the photometric discovery data, the reconnaissance
spectroscopic observations, the photometric follow-up, and the high
resolution and high signal-to-noise (S/N) spectroscopy. Then we
determine the parameters of the host star \hatcur\ (\hatcurCCgsc) in
\refsec{stelparam} by exploring a number of alternate ways. In
\refsec{blend} we investigate whether the observational data is due to
a system that mimics planetary transits, and prove that this is not the
situation, and \hatcurb\ is a {\em bona-fide} planet. We go on in
\refsec{globmod} to perform global modeling of the data to determine
system parameters, such as the orbital and planetary parameters. We
elaborate on treating the systematic variations in an optimal way, and
present a full analysis that takes them into account (\refsec{modsys},
\refsec{modphys} and the Appendix). Finally, we discuss the
implications of our findings in the Discussion (\refsec{disc}).


\section{Observations}
\label{sec:obs}
\subsection{Photometric Discovery}
\label{sec:disco}

\begin{figure}[!ht]
\plotone{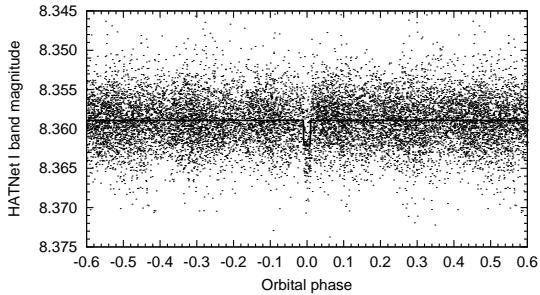}
\caption{
	The HATNet discovery \lc{} of \hatcur{} exhibiting 11470 individual
	measurements at 5.5\,min cadence. The unbinned instrumental
	\band{I} photometry was obtained with the HAT-6 (Arizona) and HAT-9
	(Hawaii) telescopes of HATNet (see text for details), and folded
	with the period of $P = \hatcurLCPprec$~days, which is the result
	of the global fit described in \refsec{globmod}. Zero orbital
	phase corresponds to the center of the transit. Superimposed is the
	so-called ``P1P3'' analytic model that was used to describe the
	HATNet data (an approximation of the \citet{mandel:2002} analytic
	formulae; see \refsec{modphys}).
\label{fig:hatnetlc}}
\end{figure}

The $8.4\arcdeg\times8.4\arcdeg$ region around \hatcurCCgsc{}, a field
internally labeled as ``G155'', was observed on a nightly basis in two
seasons, whenever weather conditions permitted. First, during the Fall
of 2004, we acquired 1213 frames with the HAT-6 instrument located at
FLWO, and 4091 frames with HAT-9 located at SMA, Mauna Kea. We
revisited the field in 2005, and acquired an additional 6166 frames
with HAT-9. Altogether we gathered 11470 5-min exposures at a
5.5-minute cadence. This unusually rich dataset was motivated by the
overlap of G155 with the field of view of the future Kepler mission.

The calibration of the HATNet frames was done utilizing standard
procedures based on IRAF\footnote{
IRAF is distributed by the National Optical Astronomy Observatories,
which are operated by the Association of Universities for Research in
Astronomy, Inc., under cooperative agreement with the National Science
Foundation.
}. The calibrated frames were then subjected to star detection and
astrometry, as described by \cite{pal:2006}. Aperture photometry using
three apertures was performed on each image at the fixed positions of
the stellar centroids, as derived from the 2MASS catalog
\citep{skrutskie:2006} and the individual astrometric solutions
relating the pixel coordinates to the world coordinate (ICRS) system of
2MASS. We extracted photometry for all 125,000 stars down to
$I\lesssim14$ that fell in the field. The raw instrumental magnitudes
$m_r$ of each individual frame were transformed to a reference frame by
fourth order polynomials in X and Y, and first order in color. The
fitted magnitudes $m_f$ yielded by the above smooth fit are used for
generating the time versus $m_f$ \lcs.

These \lcs\ have a noise characteristic that is sometimes referred to
as pink-noise \citep{pont:2006}, because it is a combination of
Gaussian white noise (due to e.g.~photon noise) and a correlated
red-noise (due to e.g.~spatial drift of stars and uncorrected flatfield
effects). The red-noise is often referred to as trends or systematic
variations. Trends in the \lcs\ have an adverse effect on detecting
shallow transiting signals: they mask the real signal, and when
phase-folded with a trial period they can also mimic transit signals.
Furthermore, for photometric follow-up data, the trends can lead to
large systematic errors in the transit parameter determination, such as
in the impact parameter or the depth of the transit. Thus, proper
treatment of systematic variations is crucial for both discovery and
accurate follow-up characterization, especially for the shallow
({\hatcurLCdip} mmag) transit of a hot Neptune presented in this paper.
The two basic methods we have employed are the External Parameter
Decorrelation (EPD) technique, briefly described earlier in
\cite{bakos:2007b}, and the Trend Filtering Algorithm
\citep[TFA;][]{kovacs:2005}. The technical details and some new
definitions are given in the Appendix (\refsec{sysvar}).

The HATNet \lcs\ were decorrelated against trends using the EPD
technique (in {\em constant EPD} mode), and subsequently by a simple
global TFA (without reconstruction, and separately from the EPD). For
the brightest stars in the field we achieved a photometric precision of
2.9\,mmag at 5.5\,min cadence. The \lcs{} were searched for periodic
box-like signals using the Box Least Squares transit-search method
\citep[BLS, see][]{kovacs:2002}. The BLS frequency spectrum of
\hatcurCCgsc{} (also known as \hatcurCCtwomass{}; $\alpha =
\hatcurCCra$, $\delta = \hatcurCCdec$; J2000) showed a number of
significant peaks, the highest one at $0.03148$\,[c/d] (31.76\,d
period), and the second one at $\sim 0.2046$\,[c/d]
($\hatcurLCPshort$\,days).  Fortunately, we inspected the second peak,
and found the corresponding transit (\reffig{hatnetlc}) with a depth of
\hatcurLCdip\,mmag, worthy of follow-up. The dip had a relative
duration (first to last contact) of $q\approx 0.0206$, equivalent to a
total duration of $Pq\approx2.41$\,hours. The signal was confirmed with
subsequent reconstructive TFA using a trapeze-shaped model function,
being the most significant periodicity in the data (the 31.76\,d signal
was probably due to a left-over systematic). The rms of the TFA
processed \lc\ was $\sim 3.2$\,mmag, making the detection of the
$\hatcurLCdip$\,mmag transit dip rather challenging, but still robust
with S/N = $11.4$. To our best knowledge, this is the shallowest
transit signal found by ground-based transit searches that belongs to a confirmed
planet. \hd{149026}, with an even shallower transit signal
\citep{winn:2008} was first detected via spectroscopy
\citep{sato:2005}. The second shallowest planetary transit found by the
transit search method is HAT-P-2b \citep{bakos:2007b}. We note that on
occasion HATNet has found even shallower transiting signals (down to
2\,mmag depth) where the transit was real, but was due to a blended
system.

\subsection{Follow-up reconnaissance spectroscopy}
\label{sec:ds}

Following the procedures described in \citet{latham:2008}, we used the
CfA Digital Speedometers \citep[DS;][]{latham:1992}, mounted on the
1.5-m Wyeth Reflector at the Oak Ridge Observatory in Harvard,
Massachusetts, and on the 1.5-m Tillinghast Reflector at FLWO, to
obtain low S/N ratio high-resolution spectra. Altogether 5
spectra were obtained between 2001 December 17 and 2007 June 5 (the
first observations were obtained for a different project to survey
dwarfs in the solar neighborhood that might be suitable targets for
SETI).  The signal-to-noise ratios ranged from 10 to 18 per spectral
resolution element of 8.5\,\kms.

Reconnaissance spectroscopy is an important step in weeding out
astrophysical false positive systems that mimic planetary transits.
Stellar parameter determination can distinguish between dwarf and giant
stars via the measurement of surface gravity \loggstar, and thus
eliminate systems where the \lc\ of an eclipsing binary is blended with
a giant star, or where the transit signal seen in our data may not be
real, as it is physically not feasible to orbit around (and outside)
the giant star with such a short period. In addition, large RV
variations (of the order of several \kms) are indicative of orbital
motion due to stellar companions rather than planets orbiting a star.
Fine analysis of the spectra can also reveal stellar triple systems.
Finally, the rapid rotation of a host star, as may be indicated by the
rotational broadening of the spectra, is often correlated with a
stellar companion that is massive enough to synchronize the rotation
with the orbital motion.

\hatcur\ survived all these steps, and the RV measurements showed an
rms residual of 0.29\,\kms, consistent with no detectable RV variation.
Initial atmospheric parameters for the star, including the effective
temperature $\teffstar = 4750\pm125$\,K, surface gravity $\loggstar =
4.5\pm0.25$ (cgs), and projected rotational velocity $\vsini = 0.0\,
\kms$, were derived as described by \citet{latham:2008}. The mean
line-of-sight velocity of the star was $\Gamma = -63.56\pm0.29\,\kms$
on an absolute scale.

\subsection{Photometric follow-up observations}
\label{sec:phfu}

Photometric follow-up observations are important  i) to perform
independent confirmation of the initial detection, ii) to allow for
accurate characterization of the system (see also \refsec{globmod}),
iii) to help eliminate blend scenarios (\refsec{blend}), and, if
multiple events are observed, iv) to possibly average out extra
variations of the \lc\ due to spots.  Since the transit of \hatcur\ was
originally detected with the 11cm diameter HATNet telescopes (although
when phase-folding many events), it is certainly feasible to confirm a
single transit with a 1m-class telescope. However, acquiring a high-quality
observation of the \hatcurLCdip\,mmag transit dip in the light of a
star as bright as $V=\hatcurCCmag$ is quite challenging. An accurate
and highly precise \lc\ is essential for eliminating blend scenarios,
and then for determining the physical parameters of the transiting
planet--star system. In general, the shallower the transit, the wider
the range of possible blends (i.e.~more than two body systems) that can
mimic the observed transit. Some of these can only be distinguished by
subtle effects, such as the duration of ingress/egress, and accurate
depth and shape of the transit.

\begin{deluxetable}{lrrrrrr}
\tablewidth{0pc}
\tablecaption{Photometric follow-up observations
	of \hatcur{}\label{tab:phfu}}
\tablehead{
	\colhead{Local Date} &
	\colhead{Instrument} &
	\colhead{Filter} &
	\colhead{$N_{tr}$} &
	\colhead{Type}
}
\startdata
    2007.0902  & FLWO12 & $z$ & $0$ & {\tt OIBEO} \\
    2007.0907  & FLWO12 & $z$ & $1$ & {\tt --BEO} \\
    2007.1021  & FLWO12 & $z$ & $10$ & {\tt OIBE(O)} \\
    2008.0518  & FLWO12 & $z$ & $53$ & {\tt (OI)BEO} \\
    2008.0603* & Schmidt & $I_C$ & $56$ & {\tt (OIB--)} \\
    2008.0701  & FLWO12 & $z$ &  $62$ & {\tt OIB--} \\
    2008.0829* & Schmidt & $I_C$ & $74$ & {\tt (OIBEO)} \\
    2008.0903  & Schmidt & $I_C$ & $75$ & {\tt OIBEO} \\
    2008.1002  & FLWO12 & $z$  & $81$ & {\tt OIBEO} \\
    2008.1007  & FLWO12 & $z$ & $82$ & {\tt OIBEO} \\
    2008.1115  & FLWO12 & $z$ & $90$ & {\tt OIB--} \\
    2008.1120  & FLWO12 & $r$ & $91$ & {\tt OIBEO}
\enddata
\tablecomments{
	Dates marked with ``*'' were of poor quality, and not used in the
	analysis. $N_{tr}$ gives the transit event number, counted from the
	0th follow-up event on 2007 September 2. The Type column shows
	which parts of the transit were caught: Out-of-transit (OOT),
	Ingress, Bottom, Egress, OOT. Parentheses mark marginal data
	quality.
}
\end{deluxetable}

\begin{figure}[!ht]
\plotone{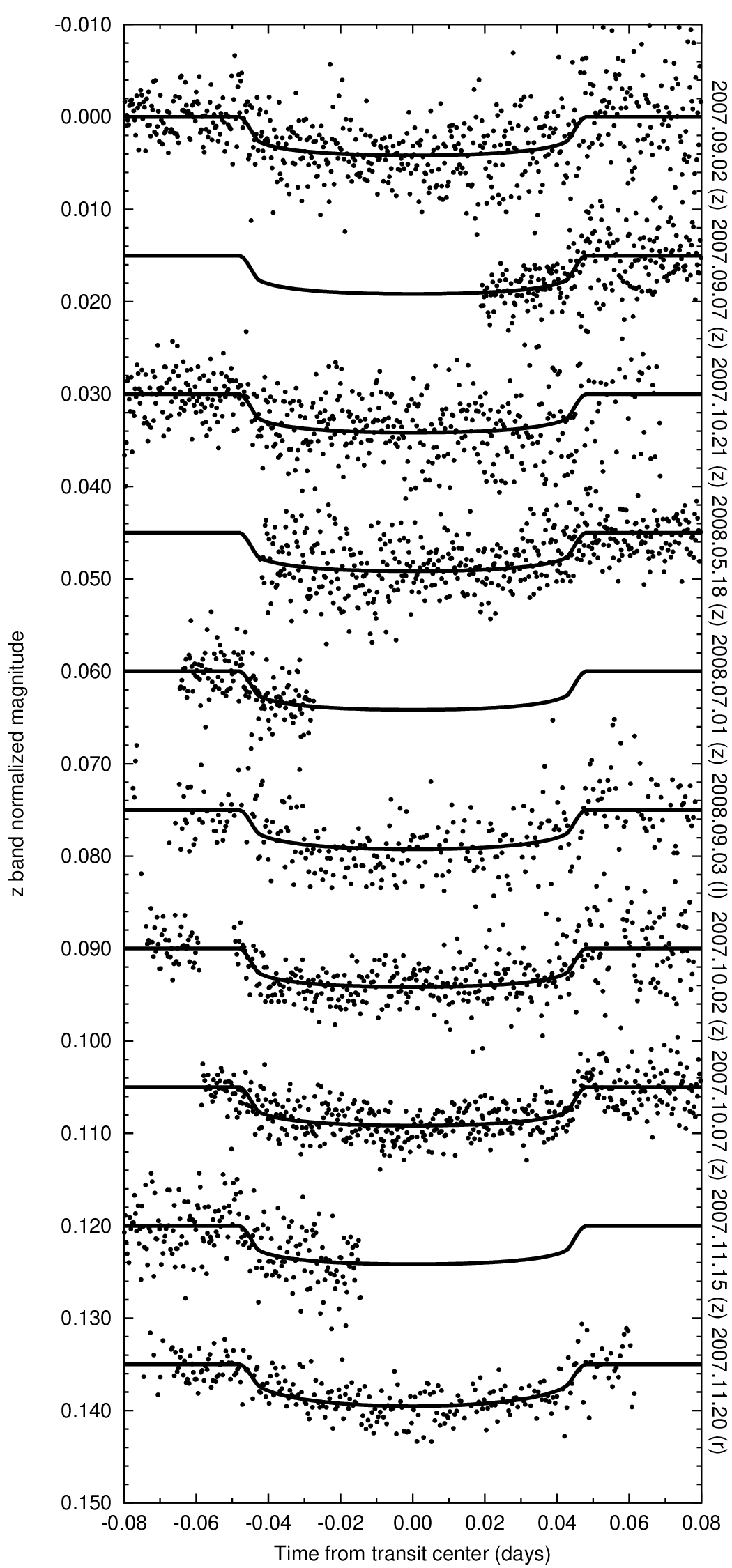}
\caption{
	Unbinned instrumental transit photometry follow-up \lcs\ acquired
	by the KeplerCam at the \flwof{} telescope (in Sloan $z$ band, and
	Sloan $r$ band) and by the 0.6\,m Schmidt telescope at
	Piszk\'estet\H{o}, Hungary (in $I_C$ band). Superimposed is the
	best-fit transit \lc{} model, yielded by the global analysis
	described in \refsec{globmod}. The \lcs\ have been subject to
	global EPD and TFA corrections using an analytical transit model
	for the intrinsic signal (see \refsec{sysvar}).
\label{fig:fuall}}
\end{figure}


For these reasons, we launched an extensive photometry follow-up
campaign, and attempted transit observations of \hatcurb\ altogether 12
times in 2007 and 2008, leading to $\sim10$ successful observations of
partial or full transits (see \reftab{phfu}). We primarily used the
\flwof\ telescope and the KeplerCam CCD in Sloan $z$-band, with
exposure times of $\sim 10$\,sec and read-out time of $12$\,sec. The
last observation on 2008 November 20/21 MST was taken through Sloan $r$
band to complement our blend analysis (\refsec{blend}). We also
observed transits 3 times using the 0.6\,m Schmidt telescope of Konkoly
Observatory at the Piszk\'estet\H{o} mountain station. These
observations were taken through \band{I_C}, and the only useful dataset
proved to be the one from 2008 September 3/4 CET.

Data were reduced in a similar manner as for the HATNet data
(\refsec{disco}), and as described in \citet{bakos:2008}. Following
bias and flat calibration, we derived an initial first order
astrometrical transformation between the $\sim750$ brightest stars and
the 2MASS catalog. The position of \hatcur\ in the 2MASS catalog was
corrected for each observed epoch before deriving the astrometry,
due to its moderately high proper motion
\citep[\hatcurCCpm\,\masyr;][]{perryman:1997}. Photometry was carried
out for all stars in the field using three apertures that were adjusted
each night to match the observing conditions (sky background, profile
width). For the nights with good seeing, one aperture was kept small to
avoid the faint neighbor 2MASS 19505049+4805017 currently at approx
6\arcsec\ distance (see \refsec{blend}). Instrumental magnitudes were
transformed to a photometric reference frame (selected to be at low
airmass, low sky background, etc.) using a first order polynomial in
the X, Y pixel coordinates and the 2MASS J-K color of $\sim 600$ stars.
The transformation was iteratively determined by weighting with
individual Poisson noise errors of the stars, and rejecting 3-$\sigma$
outliers, plus eliminating the main target and its faint neighbor from
the fit. The smooth fit was repeated based on the $\sim 180$ best
stars, using the median of the individual \lc\ magnitude values as a
new reference system, and weighting the fit with the rms of the \lcs\
(i.e.~substituting the former Poisson errors).

The \flwof\ KeplerCam observations usually result in high quality
photometry of $\sim$1\% deep planetary transits \citep[see
e.g.][]{latham:2008,bakos:2008}, because of the large field-of-view
(FOV = \fovsize{23}{\arcmin}) with many potential comparison stars, fine
pixel resolution (0.336\pxs), good quality and high quantum efficiency
CCD (monolithic, \ccdsize{4K} Fairchild 486 chip), good sky conditions from
FLWO, the fast readout through 4 channels, and careful data processing.
The performance on this very shallow transit of a fairly bright star,
however, turned out to be slightly sub-optimal with some residual
trends, necessitating diverse application of the EPD and TFA methods.
Since these were part of our global modeling of the data, including the
HATNet and RV data, they are detailed later in \refsec{globmod} and in
the Appendix. The follow-up \lcs\ after applying simultaneous EPD-TFA
with per-night coefficients for EPD and global coefficients for TFA are
displayed on \reffig{fuall}. The model function (denoted as
$m_0(\vec{p},t_i)$ in \refeq{lc} in the Appendix) in the simultaneous
EPD-TFA fit is an analytical transit model from \citet{mandel:2002}.

\ifthenelse{\boolean{emulateapj}}{
        \begin{deluxetable*}{rllllll}
}{
        \begin{deluxetable}{rllllll}
}
\tablewidth{0pc}
\tablecaption{Follow-up Photometry for \hatcur}
\label{tab:phfudata}
\tablehead{
	\colhead{BJD} &
	\colhead{Mag} &
	\colhead{Error} &
	\colhead{Raw Mag} &
	\colhead{Filter} &
	\colhead{$N_{tr}$} &
	\colhead{Instrument}
}
\startdata
2454346.61746        &   -0.00134 &    0.00064 &   -0.00082 &          z &          0 &     FLWO12\\
2454346.61771        &   -0.00228 &    0.00064 &   -0.00234 &          z &          0 &     FLWO12\\
2454351.74387        &    0.00545 &    0.00066 &    0.00615 &          z &          1 &     FLWO12\\
2454351.74414        &    0.00320 &    0.00066 &    0.00134 &          z &          1 &     FLWO12\\
2454395.58419        &   -0.00509 &    0.00068 &   -0.00078 &          z &         10 &     FLWO12\\
2454395.58575        &   -0.00042 &    0.00062 &   -0.00170 &          z &         10 &     FLWO12\\
2454605.84905        &    0.00287 &    0.00065 &    0.00393 &          z &         53 &     FLWO12\\
2454605.84933        &    0.00280 &    0.00065 &    0.00763 &          z &         53 &     FLWO12\\
2454649.81716        &    0.00186 &    0.00069 &    0.00104 &          z &         62 &     FLWO12\\
2454649.81742        &    0.00168 &    0.00069 &   -0.00059 &          z &         62 &     FLWO12\\
2454713.34579        &   -0.00234 &    0.00096 &    0.00018 &          I &         75 &    Schmidt\\
2454713.34612        &   -0.00136 &    0.00099 &    0.00131 &          I &         75 &    Schmidt\\
2454742.67630        &   -0.00239 &    0.00053 &   -0.00358 &          z &         81 &     FLWO12\\
2454742.67664        &   -0.00049 &    0.00053 &   -0.00093 &          z &         81 &     FLWO12\\
2454747.57949        &    0.00194 &    0.00054 &   -0.00019 &          z &         82 &     FLWO12\\
2454747.57981        &   -0.00251 &    0.00054 &   -0.00376 &          z &         82 &     FLWO12\\
2454786.57057        &    0.00386 &    0.00061 &    0.00543 &          z &         90 &     FLWO12\\
2454786.57106        &    0.00042 &    0.00061 &    0.00307 &          z &         90 &     FLWO12\\
2454791.55380        &    0.00134 &    0.00050 &    0.00048 &          r &         91 &     FLWO12\\
2454791.55562        &   -0.00340 &    0.00050 &   -0.00356 &          r &         91 &     FLWO12\\

\enddata
\tablecomments{
column (1): Barycentric Julian Day, \\
column (2): Best detrended magnitude, normalized to 0.0 out of transit, \\
column (3): Estimated error in the best magnitude.\\
column (4): Magnitude before detrending (denoted as $m_f$ in the text).\\
column (6): Transit number. See \reftab{phfu}.\\
This table is presented in its entirety in the electronic edition
of the Astrophysical Journal. A portion is shown here for guidance
regarding its form and content.}
\ifthenelse{\boolean{emulateapj}}{
        \end{deluxetable*}
}{
        \end{deluxetable}
}


\subsection{High resolution, high S/N spectroscopy}
\label{sec:hires}

We started observations of \hatcur\ on 2007 August 22 with the HIRES
instrument \citep{vogt:1994} on the Keck-I telescope located on Mauna
Kea, Hawaii. It was soon realized that the target is unusually
complicated because of the small RV amplitude with respect to the
moderate velocity jitter due to the active K dwarf star, and also
because of long term trend(s) present. Thus, \hatcur\ has been
extensively observed over the past two years, and we have gathered
altogether 50 spectra and 3 template observations (see \reftab{rvs}).
This is 5--10 times the number of spectra collected for a typical
HATNet transit candidate.

The width of the spectrometer slit used on HIRES was $0\farcs86$,
resulting in a resolving power of $\lambda/\Delta\lambda \approx
55,\!000$, with a wavelength coverage of $\sim3800-8000$\,\AA\@. The
iodine gas absorption cell was used to superimpose a dense forest of
$\mathrm{I}_2$ lines on the stellar spectrum and establish an accurate
wavelength fiducial \citep[see][]{marcy:1992}. Relative RVs in the
Solar System barycentric frame were derived as described by
\cite{butler:1996}, incorporating full modeling of the spatial and
temporal variations of the instrumental profile. The final RV data and
their errors as a function of barycentric Julian date are listed in
\reftab{rvs}.  It is reassuring that a simple Fourier analysis of the
RV data {\em without} prior assumption on any periodic event yields a
primary period of $4.888\pm0.019$\,days, confirming the same periodic
phenomenon that is present in the independent dataset consisting of the
discovery and follow-up photometry.\footnote{The uncertainty of the
period was simply derived from the full width at half magnitude of the
appropriate peak in the spectrum.}

Based on the numerous observations of a transit-like event in the
photometry, our initial physical model was thus a single body orbiting
and transiting the star, causing both the RV variations and the
transits. The RV variations induced on the parent star can be
characterized by six parameters: the period $P$, the center of transit
$T_c$ (i.e.~the phase of the RV curve), the semi-amplitude $K$, the RV
zero-point $\gamma$, and the Lagrangian orbital elements
$(k,h)=e\times(\cos\omega,\sin\omega)$. An orbital fit to the RV data
without any constraints from the photometry yielded the following
values:
$P=4.8896\pm0.0017$\,d,
$T_c=2454552.04\pm0.16$\,(BJD),
$K=13.1\pm2.1\,\mathrm{m\,s^{-1}}$,
$\gamma=0.9\pm0.8\,\mathrm{m\,s^{-1}}$ (arbitrary scale),
$k=0.28\pm0.12$ and
$h=0.09\pm0.09$. 

Later (\refsec{globmod}) we present a global modeling of the data,
where joint analysis of the photometry and RV data is performed. This
yields precise $P$ and $T_c$ parameters that are almost entirely
constrained by the photometric data (notably the sharp ingress/egress
features), with virtually no coupling from the RV data. The knowledge
of period and phase from photometry is a very tight constraint in the
analysis of the RV data, yielding significant improvement on
determination of the other orbital parameters. It is thus well
justified to analyze the RV data separately by fixing the ephemeris to
that determined by the photometry, and assuming that there is an
eccentric orbital motion present. In addition, we can also investigate
whether our initial model correctly described the physics of the
system, and whether there are other signals present in the RV data.

\begin{figure} 
\plotone{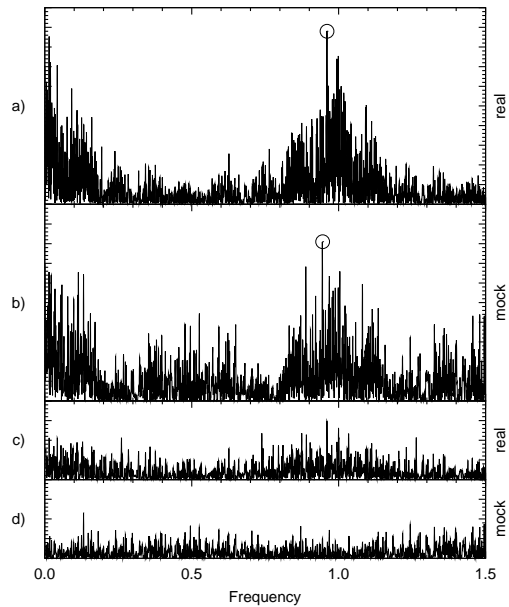}
\caption{
	(Top panel:) The inverse \chisq\ as a function of $f_2$ frequency
	when fitting a combined Keplerian orbit with fixed $P$ and $T_c$
	ephemeris, plus a sinusoidal component with $f_2$ frequency to the RV
	observations. The high peaks indicate good fit with small \chisq.
	The $0.96 [c/d]$ peak is marked with a circle.
	(Second panel:) The same plot, but on a mock dataset that consists of
	a Keplerian, a drift, and Gaussian noise. 
	(Third panel:) Same as the top panel, but the fitted model was
	extended by a drift with fitted slope. Much of the 1\,[c/d]
	frequencies disappeared, but the one at e.g.~0.9602\,[c/d]
	remained. 
	(Bottom panel:) Same as the second panel (mock data), with fitted
	model extended with a linear slope.
\label{fig:rvfreq}}
\end{figure}

Along these lines, our refined model for the RV data was an eccentric
orbital motion with $P$ and $T_c$ fixed to those values found from the
global modeling (and primarily constrained by the photometry), plus a
sinusoidal motion of the form $A_2\sin(f_2 t) + B_2\sin(f_2 t)$
(equivalent to $A_2^{\prime}\sin(f_2 t + \phi)$, but linear in the
fitted $A_2$ and $B_2$ parameters).  We searched the $f_2$ domain by
fitting $\gamma$, $K$, $k$, $h$ (eccentric orbit parameters) and $A_2$,
$B_2$ at each $f_2$, and noting the \chisq. This ``frequency scan''
located a number of significant peaks with small inverse \chisq\
(\reffig{rvfreq}, upper panel), the most significant being $f_2 =
0.9602\pm0.0009$. We suspected that the peaks around 1\,[c/d] might be
due to a long-term trend in the RV, sampled with a daily and
lunar-cycle periodicity. To test this effect we generated a mock signal
by co-adding an analytic eccentric orbital motion, a long term RV
drift, and Gaussian noise, and sampled these at the exact times of the
50 Keck observations. We ran the same fitting procedure (Keplerian plus
sinusoidal motion) in the above described frequency scan mode on the
mock data. Indeed, the drift appeared as a number of significant
frequencies around 1\,[c/d] (\reffig{rvfreq}, panel b), their exact
location depending on the amplitude of the drift and the random seed
used to generate the noise. Then, we repeated the above analysis
including a long term drift in the fitted model function while
performing the frequency scanning procedure (i.e.~fitting at each
frequency value that is being stepped on a grid).

The results are plotted in the c) panel of \reffig{rvfreq}, on the same
scale as used in panel a) and b). It can be clearly seen that the
magnitude of the residuals is definitely smaller and the strong
structures disappeared.  Using the mock data set, the results are
almost the same (\reffig{rvfreq} panel d), confirming our assumptions
that a long term drift in the data causes aliases around 1\,[c/d]. 
This is in line with the experience based on long-term monitoring of
stars with Keck/HIRES: in nearly every case where there is either a
planet or a long-term trend, a spurious spike appears in the
periodogram at close to 1 day.

Note, however, that in spite of the clearing of peaks around 1\,[c/d],
the $0.9602$\,[c/d] peak remained present even after the
simultaneous fit of a Keplerian orbit and a drift to the Keck data. 

It is also interesting that the \chisq\ of the fit with Keplerian plus
$f_2$ sinusoid component is somewhat smaller than the fit with
Keplerian plus long term trend. The $f_2 = 0.9602$\,[c/d] periodicity
may be a real physical effect, or can be an alias of a real physical
effect with different period (rotation period, activity of star), or an
alias of a systematic effect (lunar cycle, instrumental effect).
Altogether we have three components to consider in the final model of
the RV variations: i) Keplerian orbit, ii) long-term drift, iii) $f_2 =
0.9602$\,[c/d] periodicity.

\begin{figure}  
\plotone{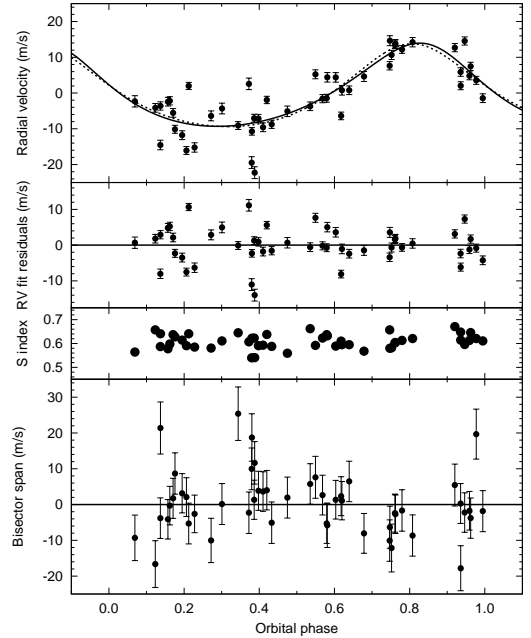}
\caption{
	(Top panel:) The radial velocity measurements from Keck
	for \hatcur{} after correction for a linear drift seen in the data. 
	The measurements are folded using our best-fit ephemeris from
	global modeling of the photometry and RV data (see
	\refsec{globmod}), and superimposed by our (eccentric) orbital fit. 
	The center-of-mass velocity has been subtracted. The orbital phase
	is shifted to be zero at the center of the transit. The error-bars
	have not been inflated with the jitter ($5.01$\,\ms). The dashed
	line (barely discernible from the thick line) is the orbital fit
	with the same ephemeris and $K$ semi-amplitude, 
	but using the refined $k_C$ and $h_C$ Lagrangian orbital elements
	(\refsec{kh}).
	(Second panel:) RV residuals after the orbital fit. 
	(Third panel:) The S activity index of \hatcur\ phase-folded.
	(Bottom panel:) Bisector spans (BS) for the Keck spectra plus the
	three template spectra, computed as described in the text. The
	mean value has been subtracted. The vertical scales on the lower
	and upper panels are the same.
\label{fig:rvbis}}
\end{figure}

The model of a Keplerian orbit and a long-term drift (i+ii) has
simple underlying physics, by assuming an inner and an outer planet.
The results of the orbital fit for this basic model are exhibited on
\reffig{rvbis}. The data folded with the $P=\hatcurLCPprec$\,d period
after removal of the best-fit drift is plotted in the top panel.
Superimposed is the Keplerian orbit that is clearly eccentric. The
orbital elements for this fit were:
$K=\hatcurRVK$\,\ms,
$k=e\cos\omega=\hatcurRVk$,
$h=e\sin\omega=\hatcurRVh$,
$\gamma=\hatcurRVgamma$\,\ms, and the
best fit value for the linear drift is 
$G_1 = \hatcurRVlindrift\,\ms\,day^{-1}$.
Note that the $K$ amplitude of the orbit is only $\hatcurRVK$\,\ms.
In order
to have a reduced \chisq\ value of $1.0$, a jitter of $5.01$\,\ms\ has to
be added to the formal errors. 

We also derived an orbital fit with a full model of all three
components (Keplerian, drift, short-period sinusoidal) so as to check
the effect of $f_2$ on the orbital parameters. We found that the change
in the orbital elements is insignificant, with small decrease
in their respective error-bars:
$K=10.91\pm0.96$\,\ms,
$k=e\cos\omega=0.225\pm0.038$,
$h=e\sin\omega=0.069\pm0.072$,
drift $G_1 = 0.0176 \pm 0.0056\,\ms\,day^{-1}$,
and the amplitude of the $f_2$ sinusoid is $5.7\pm1.3$\,\ms. The jitter
value changed to $3.89$\,\ms.  

We also investigated models with non-linear drift, as characterized by
$G_2$ quadratic and $G_3$ cubic terms, and found that these are
insignificant based on the present data. We also checked for
correlations between the RV residuals from the best fit, and the
spectral bisector-spans, and the $S$ activity index, but found that
these correlations were insignificant.

Given the fact that the origin of the $f_2$ periodicity (component iii)
is unknown, and it is suspected to be an alias that may diminish by
taking more data, plus the key orbital elements do not change
significantly by taking it into account, in the rest of this paper we
adopted the simpler model of a Keplerian orbit (with
$K=\hatcurRVK\,\ms$) and a long-term drift {\em without} the short-term
periodicity.

\ifthenelse{\boolean{emulateapj}}{
    \begin{deluxetable*}{lcrrrrrrr}
}{
    \begin{deluxetable}{lcrrrrrrr}
}
\tablewidth{0pc}
\tablecaption{Relative radial velocity measurements 
	of \hatcur{}\label{tab:rvs}}
\tablehead{
	\colhead{BJD} &
	\colhead{RV} &
	\colhead{\ensuremath{\sigma_{\rm RV}}} &
	\colhead{O-C} &
	\colhead{BS} &
	\colhead{\ensuremath{\sigma_{\rm BS}}} &
	\colhead{S}\\
	&
	\colhead{(\ms)} &
	\colhead{(\ms)} &
	\colhead{(\ms)} &
	\colhead{(\ms)} &
	\colhead{(\ms)} &
		\\
}
\startdata
\ifthenelse{\boolean{astroph}}{
2454335.89332 \dotfill & $      8.31$ & $      1.10$ & $      1.82$ &   $  -2.40 $ & $  5.32 $   & $ 0.603  $  \\ 
2454335.89997 \dotfill & $      8.61$ & $      1.17$ & $      2.03$ &   $  -2.70 $ & $  5.38 $   & $ 0.604  $  \\ 
2454336.74875 \dotfill & $     -3.12$ & $      1.12$ & $     -6.29$ &   $ -17.80 $ & $  6.29 $   & $ 0.614  $  \\ 
2454336.25715 \dotfill &    \nodata   &     \nodata  &   \nodata    &   $  -1.30 $ & $  5.27 $   & $ 0.616  $  \\ 
2454336.86162 \dotfill & $     -0.34$ & $      1.11$ & $     -1.36$ &   $  -1.74 $ & $  5.41 $   & $ 0.613  $  \\ 
2454336.94961 \dotfill & $     -1.60$ & $      1.13$ & $     -0.93$ &   $  19.68 $ & $  6.98 $   & $ 0.621  $  \\ 
2454337.73150 \dotfill & $     -8.74$ & $      1.15$ & $      2.78$ &   $  21.40 $ & $  7.28 $   & $ 0.641  $  \\ 
2454337.92101 \dotfill & $    -15.25$ & $      1.11$ & $     -2.42$ &   $   8.67 $ & $  5.78 $   & $ 0.628  $  \\ 
2454338.74219 \dotfill & $    -14.18$ & $      1.10$ & $      0.02$ &   $  25.38 $ & $  7.47 $   & $ 0.645  $  \\ 
2454338.92046 \dotfill & $    -15.82$ & $      1.06$ & $     -2.09$ &   $  18.73 $ & $  6.63 $   & $ 0.622  $  \\ 
2454339.89292 \dotfill & $     -6.53$ & $      1.10$ & $     -0.09$ &   $  -5.27 $ & $  5.60 $   & $ 0.637  $  \\ 
2454343.83879 \dotfill & $    -11.99$ & $      1.10$ & $      1.48$ &   $   1.34 $ & $  5.47 $   & $ 0.623  $  \\ 
2454344.96843 \dotfill & $    -11.31$ & $      1.07$ & $     -7.44$ &   $   2.34 $ & $  5.44 $   & $ 0.610  $  \\ 
2454344.47643 \dotfill &   \nodata    &    \nodata   &   \nodata    &   $  -9.38 $ & $  5.69 $   & $ 0.616  $  \\ 
2454344.48214 \dotfill &   \nodata    &    \nodata   &   \nodata    &   $  -9.88 $ & $  5.57 $   & $ 0.612  $  \\ 
2454396.75318 \dotfill & $     -1.40$ & $      1.00$ & $     10.52$ &   $  -5.31 $ & $  5.69 $   & $ 0.641  $  \\ 
2454397.76902 \dotfill & $     -5.30$ & $      1.04$ & $      5.85$ &   $   4.01 $ & $  5.51 $   & $ 0.638  $  \\ 
2454427.72779 \dotfill & $      2.75$ & $      1.26$ & $      8.18$ &   $   7.61 $ & $  5.87 $   & $ 0.592  $  \\ 
2454428.69600 \dotfill & $     12.16$ & $      1.39$ & $      3.77$ &   $  -6.33 $ & $  5.88 $   & $ 0.580  $  \\ 
2454430.69738 \dotfill & $     -4.82$ & $      1.22$ & $      4.64$ &   $  -4.12 $ & $  5.52 $   & $ 0.578  $  \\ 
2454454.70749 \dotfill & $     -4.02$ & $      1.58$ & $      0.48$ &   $  -9.30 $ & $  6.36 $   & $ 0.564  $  \\ 
2454455.69785 \dotfill & $     -8.06$ & $      1.39$ & $      2.78$ &   $ -10.02 $ & $  6.20 $   & $ 0.581  $  \\ 
2454456.68976 \dotfill & $     -6.75$ & $      1.47$ & $      1.02$ &   $   1.94 $ & $  5.72 $   & $ 0.559  $  \\ 
2454457.68670 \dotfill & $      3.02$ & $      1.39$ & $     -0.98$ &   $  -8.06 $ & $  5.55 $   & $ 0.568  $  \\ 
2454460.72727 \dotfill & $     -5.84$ & $      1.51$ & $      4.91$ &   $   0.15 $ & $  5.72 $   & $ 0.611  $  \\ 
2454549.09456 \dotfill & $    -18.39$ & $      1.66$ & $    -10.98$ &   $   9.97 $ & $  5.83 $   & $ 0.540  $  \\ 
2454549.13273 \dotfill & $    -21.19$ & $      1.65$ & $    -13.92$ &   $  11.64 $ & $  5.93 $   & $ 0.541  $  \\ 
2454602.01197 \dotfill & $    -13.45$ & $      1.10$ & $     -7.82$ &   $   2.08 $ & $  5.42 $   & $ 0.591  $  \\ 
2454602.11819 \dotfill & $    -12.58$ & $      1.31$ & $     -6.57$ &   $  -2.61 $ & $  5.25 $   & $ 0.585  $  \\ 
2454602.94861 \dotfill & $     -4.49$ & $      1.19$ & $      0.97$ &   $   3.85 $ & $  5.46 $   & $ 0.591  $  \\ 
2454603.00783 \dotfill & $     -6.92$ & $      1.23$ & $     -1.71$ &   $   3.64 $ & $  5.54 $   & $ 0.593  $  \\ 
2454603.11953 \dotfill & $     -6.07$ & $      1.16$ & $     -1.42$ &   $  -5.07 $ & $  5.78 $   & $ 0.588  $  \\ 
2454603.95329 \dotfill & $      7.05$ & $      1.34$ & $      4.05$ &   $   1.34 $ & $  5.40 $   & $ 0.589  $  \\ 
2454604.03295 \dotfill & $      3.47$ & $      1.35$ & $     -0.60$ &   $   1.08 $ & $  5.37 $   & $ 0.595  $  \\ 
2454604.12769 \dotfill & $      3.48$ & $      1.21$ & $     -1.95$ &   $   6.46 $ & $  5.65 $   & $ 0.595  $  \\ 
2454634.00592 \dotfill & $     14.22$ & $      1.27$ & $     -0.66$ &   $ -12.14 $ & $  6.68 $   & $ 0.582  $  \\ 
2454634.95641 \dotfill & $     18.14$ & $      1.14$ & $      6.98$ &   $  -2.23 $ & $  5.52 $   & $ 0.596  $  \\ 
2454636.01011 \dotfill & $      1.45$ & $      1.16$ & $      4.90$ &   $  -0.31 $ & $  5.39 $   & $ 0.598  $  \\ 
2454637.99531 \dotfill & $      2.11$ & $      1.13$ & $      0.21$ &   $   2.64 $ & $  5.55 $   & $ 0.623  $  \\ 
2454639.02952 \dotfill & $     15.89$ & $      1.08$ & $     -0.81$ &   $  -1.66 $ & $  5.79 $   & $ 0.613  $  \\ 
2454640.07995 \dotfill & $      2.31$ & $      1.22$ & $     -4.53$ &   $  -1.84 $ & $  5.73 $   & $ 0.611  $  \\ 
2454641.05878 \dotfill & $     -8.01$ & $      1.24$ & $     -3.80$ &   $   3.15 $ & $  5.49 $   & $ 0.614  $  \\ 
2454674.98972 \dotfill & $     -9.75$ & $      1.33$ & $     -8.38$ &   $  -3.81 $ & $  5.67 $   & $ 0.587  $  \\ 
2454722.80946 \dotfill & $     18.87$ & $      1.16$ & $      2.70$ &   $   5.44 $ & $  5.86 $   & $ 0.671  $  \\ 
2454723.80199 \dotfill & $      2.07$ & $      1.21$ & $      1.40$ &   $ -16.64 $ & $  6.54 $   & $ 0.657  $  \\ 
2454725.81649 \dotfill & $      2.56$ & $      1.21$ & $     -0.29$ &   $   5.75 $ & $  5.67 $   & $ 0.662  $  \\ 
2454726.85018 \dotfill & $     13.91$ & $      1.17$ & $     -3.43$ &   $ -10.07 $ & $  5.78 $   & $ 0.657  $  \\ 
2454727.77201 \dotfill & $     12.20$ & $      1.20$ & $     -2.75$ &   $   0.29 $ & $  5.57 $   & $ 0.648  $  \\ 
2454727.90334 \dotfill & $     13.77$ & $      1.16$ & $      1.32$ &   $  -3.79 $ & $  5.63 $   & $ 0.646  $  \\ 
2454777.79897 \dotfill & $      2.23$ & $      1.21$ & $      1.68$ &   $   1.74 $ & $  5.59 $   & $ 0.637  $  \\ 
2454778.78504 \dotfill & $     10.39$ & $      1.61$ & $     11.00$ &   $  -2.29 $ & $  5.80 $   & $ 0.607  $  \\ 
2454779.80416 \dotfill & $     12.24$ & $      1.33$ & $      5.34$ &   $  -5.81 $ & $  6.20 $   & $ 0.632  $  \\ 
2454790.68835 \dotfill & $     22.38$ & $      1.31$ & $      0.04$ &   $  -8.66 $ & $  5.76 $   & $ 0.621  $  \\ 

}{
2454335.89332 \dotfill & $      8.31$ & $      1.10$ & $      1.82$ &   $  -2.40 $ & $  5.32 $   & $ 0.603  $  \\ 
2454335.89997 \dotfill & $      8.61$ & $      1.17$ & $      2.03$ &   $  -2.70 $ & $  5.38 $   & $ 0.604  $  \\ 
2454336.74875 \dotfill & $     -3.12$ & $      1.12$ & $     -6.29$ &   $ -17.80 $ & $  6.29 $   & $ 0.614  $  \\ 
2454336.25715 \dotfill &    \nodata   &     \nodata  &   \nodata    &   $  -1.30 $ & $  5.27 $   & $ 0.616  $  \\ 
2454336.86162 \dotfill & $     -0.34$ & $      1.11$ & $     -1.36$ &   $  -1.74 $ & $  5.41 $   & $ 0.613  $  \\ 
\nodata       \dotfill &    \nodata   &     \nodata  &   \nodata    &   \nodata    & \nodata     & \nodata     \\ 
2454790.68835 \dotfill & $     22.38$ & $      1.31$ & $      0.04$ &   $  -8.66 $ & $  5.76 $   & $ 0.621  $  \\ 

}
\enddata
\ifthenelse{\boolean{astroph}}{
}{
	\tablecomments{
		This table is presented in its entirety in the electronic edition
		of the Astrophysical Journal. A portion is shown here for guidance
		regarding its form and content.
}
}
\ifthenelse{\boolean{emulateapj}}{
    \end{deluxetable*}
}{
    \end{deluxetable}
}


\section{Properties of the parent star}
\label{sec:stelparam}

Knowledge of the parameters of the host star is important, because it
puts the relative quantities arising from the global modeling of the
photometric and RV data (\refsec{globmod}) on an absolute scale, since
the planetary radius $\rpl$ is $\propto \rstar$ and the planetary mass
$\mpl$ is $\propto \mstar^{2/3}$. Also, modeling of the
photometric transit requires limb-darkening parameters. These may be
fitted along with other parameters if the data are of high quality and
allow it \citep{brown:2001}.  Alternatively, limb-darkening coefficients
are taken from look-up tables, such as that of \cite{claret:2004},
which depend on stellar atmospheric parameters, primarily effective
temperature. Conversely, the presence of the planetary transit imposes
constraints on the stellar parameters through the normalized semi-major
axis $a/\rstar$ (and stellar density \rhostar) as yielded by the global
modeling of the photometry, and, for eccentric orbits, the RV 
data (T08).

\subsection{Basic Stellar Parameters}
\label{sec:basicstel}

We employed the Spectroscopy Made Easy (SME) package of
\cite{valenti:1996} along with the atomic-line database of
\cite{valenti:2005} to derive an initial value for the stellar
atmospheric parameters. In this analysis we used the iodine-free
template spectrum obtained by the HIRES instrument on Keck~I. The SME
analysis of stellar spectra resulted in a stellar surface gravity
$\loggstar=4.7\pm0.1$~(CGS), metallicity $\feh = 0.32\pm0.06$\,dex,
effective temperature $\teffstar=4850\pm50$\,K, and the projected
rotational velocity $\vsini = 0.5\pm0.5$\,\kms. However, the \vsini\
value depends inversely on the assumed value of macro-turbulence
$v_{mac}$ in the star. The value of $v_{mac}$ assumed in the SME
derivation was 2.57\,\kms, but for stars of type K0 or later, $v_{mac}$
may be as small as 1.3\,\kms\ (e.g.~Gray 1988). With that value of
%
%
$v_{mac}$, \vsini\ could be as high as 2.7\,\kms. Because of the
indeterminacy of $v_{mac}$ for \hatcur, we can conclude solely that
\hatcur\ is a relatively slow rotator, and assign it a projected
rotational velocity $\vsini = 1.5\pm1.5$\,\kms, which encompasses both
extremes of \vsini\ noted here.

At this stage we could use the effective temperature as color indicator
and the surface gravity as luminosity indicator, and determine the
stellar parameters based on these two constraints using a set of
isochrones. However, it has been shown \citep{sozzetti:2007} that
\loggstar\ has a subtle effect on the spectral line shapes, and is
usually not the best luminosity indicator. For planetary transits, the
$a/\rstar$ normalized semi-major axis and related $\rhostar$ mean
stellar density typically impose a stronger constraint on possible
stellar models \citep{sozzetti:2007}. (The use of \arstar\ is discussed
later in \refsec{arstar}.) However, for \hatcur{} there is an even
better luminosity indicator, since it is a bright and nearby star with
known parallax of small relative error.
\hatcur{} enters the Hipparcos catalogue as \hip{97657} with reported
parallax of $27.50\pm0.96$\,mas \citep{perryman:1997}, equivalent to a
distance of $36.4\pm1.3$\,pc and distance modulus of
$\Delta=2.80\pm0.08$. Combination of the distance information with the
apparent brightnesses in various photometric bands yields the absolute
magnitude of the star, which is a tighter constraint on the luminosity
than the \loggstar\ or \arstar\ constraints.

As regards apparent magnitude, the TASS \citep{droege:2006} photometry
for this star is 
$V_{\rm TASS}=9.587\pm0.071$ and 
$I_{\rm TASS}=8.357\pm0.050$, yielding 
$(V-I)_{\rm TASS}=1.230\pm0.087$. 
On the other hand, the Yonsei-Yale stellar evolution models
\citep{yi:2001,demarque:2004} indicate $(V-I)_{\rm
YY,expect}=0.98\pm0.03$ for a star with $\teffstar\sim 4850$\,K
(based on the SME results). The 3-$\sigma$ inconsistency cannot be
explained by interstellar reddening, since this is a close-by star.
Thus we opted not to use the TASS photometry in this analysis.

The 2MASS catalogue \citep{skrutskie:2006} provided much better agreement.
The magnitudes reported in the 2MASS catalogue have to be converted to
the standard ESO system, in which the stellar evolution
models (specifically the YY models) specify the colors. The reported
magnitudes for this star are
$J_{\rm 2MASS}=7.608\pm0.029$, 
$H_{\rm 2MASS}=7.131\pm0.021$ and 
$K_{\rm 2MASS}=7.009\pm0.020$;
which is equivalent to $J=7.686\pm0.033$, $H=7.143\pm0.028$ and
$K=7.051\pm0.021$ in the ESO photometric system
\citep[see][]{carpenter:2001}. Thus, the converted 2MASS magnitudes
yield a color of $(J-K)=0.635\pm0.043$ that is within 1-$\sigma$ of the
expected $(J-K)_{\rm YY,expect}=0.59\pm0.02$. We thus relied on the
2MASS $K$ apparent magnitude and the parallax to derive an absolute
magnitude of $M_{\rm K}=4.18\pm0.07$. The choice of $K$ band was
motivated by the longest wavelength with smallest expected
discrepancies due to molecular lines in the spectrum of this
\hatcurYYspec\ dwarf.

In practice, the isochrone search for the best fit stellar parameters
was done in a Monte Carlo way, by assuming Gaussian uncertainties for
the Hipparcos parallax, \teffstar, \feh, the apparent 2MASS magnitudes,
and the conversion coefficients by \cite{carpenter:2001} that transform
the 2MASS magnitudes to the standard $K$ band. Lacking information,
possible correlations between some of these parameters (e.g.~\teffstar\
and \feh) were ignored. A large set ($\sim5000$)
of random $\Delta$ (distance modulus), \teffstar, \feh\ and $K$ values
were generated, and for each combination we searched the stellar
evolutionary tracks of the Yonsei-Yale models for the best fit stellar
model parameters (such as \mstar, \rstar, etc). For an unevolved K star
there was no ambiguity in the solution, i.e.~we did not enter a regime
where isochrones cross each other. Certain parameter combinations in
the Monte Carlo search did not match any isochrone. In such cases
($\sim14$\% of all trials) we skipped to the next randomly drawn
parameter set. At the end we derived the mean values and uncertainties
of the physical parameters based on their a posteriori distribution. We
also refined the stellar surface gravity. The new value
$\loggstar=4.59\pm0.03$ agrees well with the earlier SME value, and has
much smaller uncertainty confirming our previous assumptions that the
absolute magnitude for this star is a better luminosity indicator than
the surface gravity.

\begin{figure*}[!ht]
\plotone{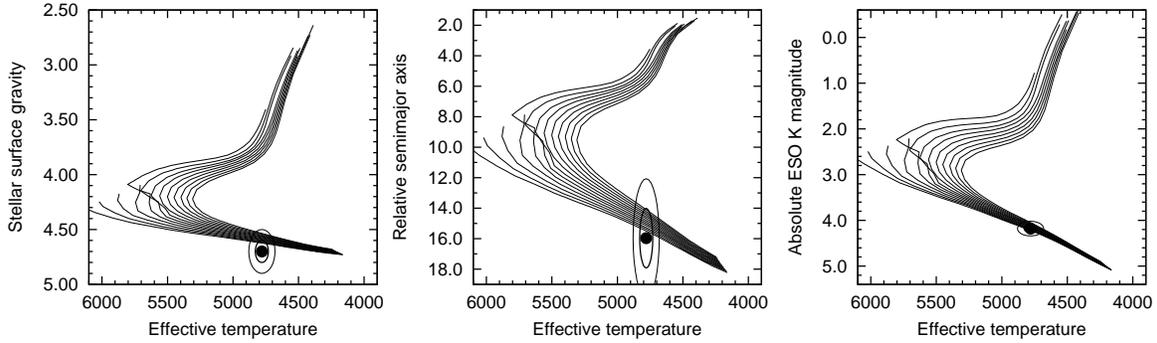}
\caption{
	The three panels display the same Yonsei-Yale isochrones between
	$2.0$ and $14.0$~Gyrs for metal-rich, $\feh=\hatcurSMEzfehNE$
	stars, including masses between $0.65\,\msun\le M \le 1.25\,\msun$. 
	The horizontal axis is effective temperature for all cases. The
	three panels show different choices of luminosity indicators (on
	the vertical axes).  The observed values for the stellar parameters
	of \hatcur\ are marked by the large filled circles, and the
	1-$\sigma$ and 2-$\sigma$ confidence ellipsoids are also indicated.
%
%
	Left panel: the luminosity indicator (vertical axis) is the stellar
	surface gravity.
	Middle panel: The luminosity indicator is the geometric semi-major
	axis \arstar, as derived from the \lc\ modeling (see
	\refsec{arstar} and \refsec{globmod}).
	Right panel: here the luminosity indicator is the $M_K$ absolute K
	magnitude, based on the 2MASS catalogue, transformations by
	\cite{carpenter:2001}, and the Hipparcos parallax.
	The tightest constraint on the isochrones is provided by the
	effective temperature (based on the SME analysis), and the absolute
	$K$ magnitude photometry (right panel). 
	\label{fig:isochrones}
}
\end{figure*}

We then repeated the SME analysis by fixing \loggstar\ to the new
value, and only adjusting \teffstar, \feh\ and \vsini. This second
iteration yielded $\teffstar=\hatcurSMEteff$\,K, and $\feh =
\hatcurSMEzfeh$. It also yielded $\vsini = 0.3\pm0.5$\,\kms, but for
the same reason as given earlier we adopt the relaxed range
$\vsini=\hatcurSMEvsin$\,\kms. We accepted the above values as the
final atmospheric parameters for this star. We then also repeated the
isochrone search for stellar parameters, yielding
\mstar=\hatcurYYmlong\,\msun, \rstar=\hatcurYYrlong\,\rsun\ and
\lstar=\hatcurYYlum\,\lsun. Along with other stellar parameters, these
are summarized in \reftab{stellar}. The stellar evolutionary isochrones
for metallicity \feh=\hatcurSMEzfehNE\ are plotted in the right panel
of \reffig{isochrones}, with the final choice of effective temperature
$\teffstar$ and the absolute magnitude $M_{\rm K}$ marked, and
encircled by the 1-$\sigma$ and 2-$\sigma$ confidence ellipsoids.

The justification for our choice of using the parallax as the
luminosity indicator is demonstrated by \reffig{isochrones}. On the
left panel we plot a set of \feh=\hatcurSMEzfehNE\ Yale isochrones as a
function of effective temperature, with the vertical axis being
\loggstar, and the observed values with their respective error
ellipsoids overlaid. The middle panel shows the same set of isochrones
with the $a/\rstar$ luminosity indicator on the vertical scale. Here
the geometric semi-major axis \arstar\ is determined from the
photometric transit and RV data (as shown later in \refsec{arstar}),
and its error is significantly increased by the uncertainties in the
eccentricity of the RV data. The right panel shows the same set of
isochrones, with vertical axis being the absolute $K$ magnitude.
Overlaid is the observational constraint based on the apparent
magnitude and Hipparcos parallax, along with the respective 1-$\sigma$
and 2-$\sigma$ error ellipsoids. It is clearly seen that the right
panel imposes the tightest constraint on the stellar parameters. We
note here that an isochrone search using the reported $J-K$ color
(instead of \teffstar\ as color indicator) also agrees with the
evolutionary models; however, the relative volume of its confidence
ellipsoid is somewhat larger.

The effective temperature from the SME analysis and the surface gravity
derived above correspond to a K4V star using \citet{gray:1992}. The
$B-V$ color index from the YY isochrones is $1.063\pm0.024$, also
consistent with the K4 spectral type of \citet{gray:1992}.  We also ran
the \citet{ramirez:2005} and \citet{casagrande:2006} temperature
calibrations in reverse to back out the $B-V$ required to produce the
SME temperature. We got $B-V = 1.025\pm 0.023$ and $1.067\pm0.025$,
respectively. For final value we accepted their average: $B-V =
1.046\pm0.024$.

\begin{deluxetable}{lcl}
\tablewidth{0pc}
\tablecaption{
	Stellar parameters for \hatcur{}
	\label{tab:stellar}
}
\tablehead{
	\colhead{Parameter}	&
	\colhead{Value} &
	\colhead{Source}
}
\startdata
$\teffstar$ (K)\dotfill         &  \hatcurSMEteff   & SME\tablenotemark{a}\\
$\feh$\dotfill                  &  \hatcurSMEzfeh   & SME                 \\
$\vsini$ (\kms)\dotfill         &  \hatcurSMEvsin   & SME                 \\
$\mstar$ ($\msun$)\dotfill      &  \hatcurYYm       & Y$^2$+Hip+SME\tablenotemark{b}\\
$\rstar$ ($\rsun$)\dotfill      &  \hatcurYYr       & Y$^2$+Hip+SME         \\
$\loggstar$ (cgs)\dotfill       &  \hatcurYYlogg    & Y$^2$+Hip+SME         \\
$\lstar$ ($\lsun$)\dotfill      &  \hatcurYYlum     & Y$^2$+Hip+SME         \\
$M_V$ (mag)\dotfill             &  \hatcurYYmv      & Y$^2$+Hip+SME         \\
Age (Gyr)\dotfill               &  \hatcurYYage     & Y$^2$+Hip+SME         \\
Distance (pc)\dotfill           &  \hatcurXdist     & Y$^2$+Hip+SME\tablenotemark{c}\\
\enddata
\tablenotetext{a}{SME = `Spectroscopy Made Easy' package for analysis
	of high-resolution spectra \cite{valenti:1996}. These parameters
	depend primarily on SME, with basically no dependence on the
	iterative analysis based on the Hipparcos parallax and YY
	isochrones (\refsec{basicstel}).}
\tablenotetext{b}{Y$^2$+Hip+SME = Yale-Yonsei 
	isochrones \citep{yi:2001}, Hipparcos distance data, and SME results.}
\tablenotetext{c}{The distance given in the table is based in the
	self-consistent analysis that relies on the Hipparcos parallax and the
	YY isochrones. It slightly differs from the Hipparcos-based distance.}
\end{deluxetable}

\subsubsection{Baraffe isochrones}
\label{sec:baraffe}

\begin{figure} 
\plotone{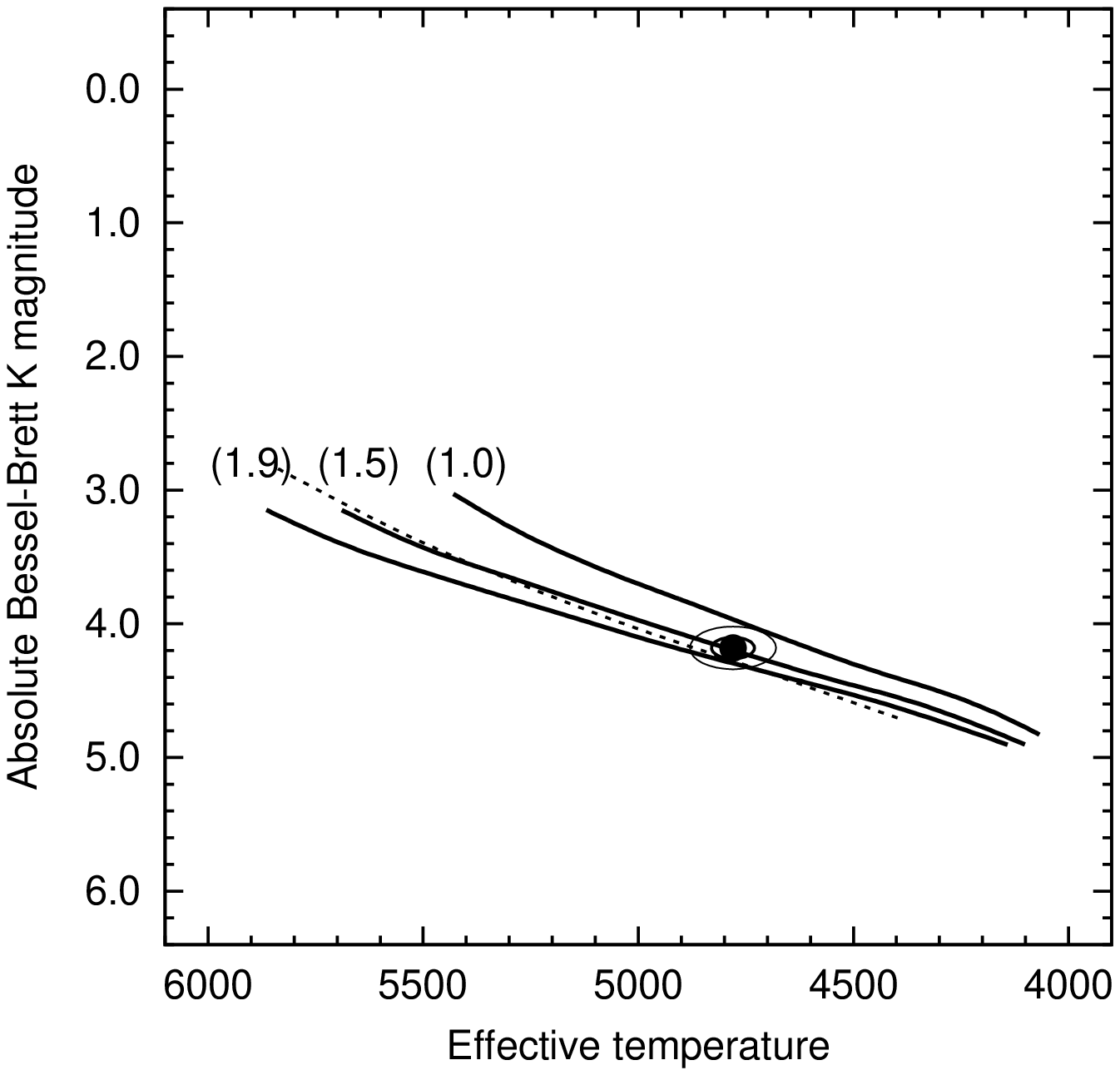}
\caption{
	\cite{baraffe:1998} isochrones with mixing length parameters
	$\alpha = 1.0$, $1.5$ and $1.9$ (solid lines) for solar metallicity
	(note that the metallicity of \hatcur\ is \hatcurSMEzfeh). A
	metal-poor ($\feh = -0.3$) isochrone for $\alpha = 1.0$ from
	\citet{baraffe:1997} is over-plotted as dashed line to investigate
	the shift due to metallicity. Approximately the opposite shift can
	be expected when increasing the metallicity to \hatcurSMEzfehNE, 
	resulting in a mis-match between the observational values and the
	isochrones.
\label{fig:baraffe}}
\end{figure}

We also investigated the dependence of stellar parameters on the choice
of isochrones. Since \hatcur\ is a K dwarf, we used the
\cite{baraffe:1998} isochrones that are usually a better choice for
late type dwarfs. \citet{baraffe:1998} presented three sets of
isochrones with different mixing length parameters. The one with
$\alpha = 1.0$ is a better match for low-mass stars (such as late K or
M dwarfs), the one with $\alpha = 1.9$ matches the Sun, and the third
set with $\alpha = 1.5$ is in between. In \reffig{baraffe} we plot
these isochrones as a function of \teffstar{}, with vertical axis being
absolute $K$ magnitude in the Bessel Brett system. Both the 2MASS $K_s$
band and the Baraffe CIT systems were transformed to $K$ using the
relations in \cite{carpenter:2001}. The Baraffe isochrones for $\alpha
> 1.0$ are given only for solar metallicity, while those for $\alpha =
1.0$ are available for solar and sub-solar \citep{baraffe:1997}
metallicity only, but not for the metal-rich composition of \hatcur{}
($\feh=\hatcurSMEzfeh$). To investigate the effect of metallicity, we
assumed the same qualitative behavior (i.e.~opposite shift in the
absolute magnitude---temperature plane) for isochrones with metallicity
increased by $+0.3$, independent of their mixing lengths. In
\reffig{baraffe}, we over-plot a metal-poor ($\feh=-0.3$, $\alpha=1.0$)
isochrone, and conclude that an opposite change in metallicity to match
that of \hatcur\ would move the $\alpha=1.0$, $\feh=+0.3$ isochrone
away from the observational values of $K$ and $\teffstar$. Because of
the lack of metal-rich models from Baraffe, and the lack of the
knowledge of the proper mixing length parameter, we omit any
quantitative conclusions from these models. Altogether, we found a
better match with the YY isochrones, and thus accepted YY-based stellar
parameters as final values (\reftab{stellar}).

\subsection{Constraints on stellar parameters by the normalized
	semi-major axis}
\label{sec:arstar}

As noted by \cite{sozzetti:2007,torres:2008}, a possible
luminosity indicator for host stars of transiting planets is the
$\arstar$ quantity, where $a$ is the relative semi-major axis, and
\rstar\ is the radius of the host star. Here \arstar\ is in simple
relation with the mean stellar density, if we assume that the planetary
mass is much smaller than the stellar mass:
\begin{equation}
	\label{eq:arstar}
	\rho_\star = 0.01892 \frac{(a/\rstar)^3}{(P/{\rm day})^2}\,\gcmc.
\end{equation}

Analysis of the transit \lc\ yields -- among other parameters -- the
quantity \zrstar\, that is related to the time spent in between the
planetary center crossing the limb of the star as $T_{dur} =
(2\zrstar)^{-1}$. For circular orbits and equatorial transits \arstar\ is
a function of \zrstar\ and the orbital period, and thus can be determined
in a straightforward way. For eccentric orbits, the relation between
\zrstar\ and \arstar, as based on \citet{murray:1999} Eq.~236
or \citet{tingley:2005}:
\begin{equation}
	\frac{\zeta}{\rstar}=\frac{a}{\rstar}\frac{2\pi}{P}\frac{1}{\sqrt{1-b^2}}
	\frac{1+h}{\sqrt{1-e^2}} .
\label{eq:zetavsar}
\end{equation}
Thus, in addition to the \lc\ parameters, \arstar\ also depends on the
orbital eccentricity and argument of pericenter, and uncertainties in
these parameters propagate into the error of \arstar.

In the case of \hatcur, the RV data show a significant eccentricity of
$e=\hatcurRVeccen$ (\refsec{globmod}). Although \zrstar\ and the
$\sqrt{1-b^2}$ term have small uncertainties (1.0\% and 4.8\%,
respectively), the uncertainty in the $(1+h)/\sqrt{1-e^2}$ term is
higher, yielding a final value of $\arstar=14.6_{-1.4}^{+1.7}$. The
significant error can be credited to the uncertainties in the orbital
parameters, as caused by the small RV amplitude and the stellar jitter.
For comparison, if \arstar\ is calculated backwards from the Hipparcos
parallax based stellar evolutionary modeling, we get $\arstar =
\hatcurPPar$. It is reassuring that this is consistent with that
derived from the global modeling of the data, but the uncertainties are
$\sim3-4$ times smaller. This justifies the choice of the
Hipparcos-parallax based luminosity indicator over using the \arstar\
constraint from global modeling of the data. This is also confirmed by
comparing \reffig{isochrones} panels b and c, where the confidence
ellipsoids from the parallax constraint (panel c) impose a tighter
constraint on the isochrones.

\subsection{Rotation and Activity of \hatcur} 
\label{sec:stelact}

\begin{figure*}[!ht]
\plotone{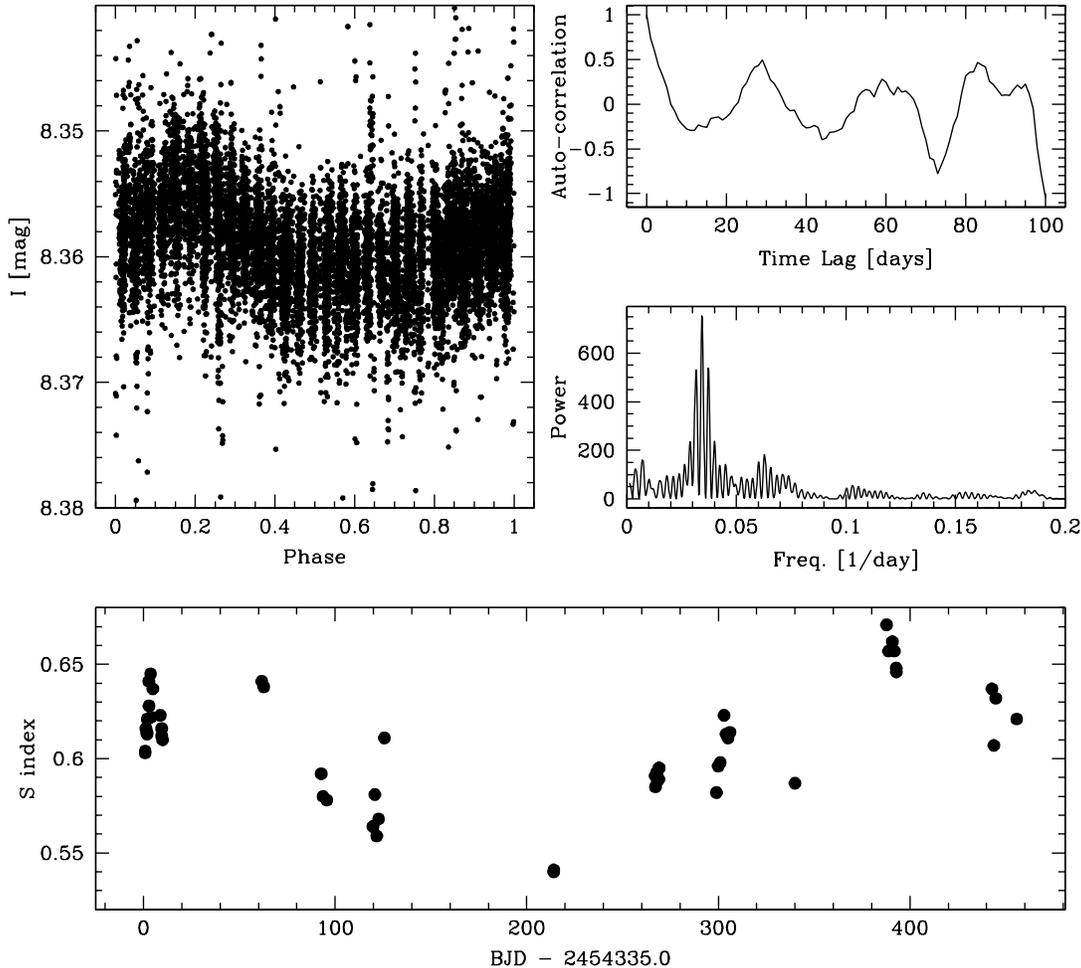}
\caption{
	(Left) The pre-TFA, post-EPD HATnet $I$-band \lc\  of \hatcur\
	phased at a period of $P=29.2$~days. Only out-of-transit points are
	included in this plot. The apparent 6.4\,mmag peak-to-peak
	variation may be due to the rotational modulation of starspots on
	the surface of \hatcur. (Upper Right) The auto-correlation function
	of the \lc\ at left. Note the first peak at a time-lag of 29~days. 
	(Middle Right) The Lomb-Scargle periodogram of the \lc\ at left,
	Note the peak at a frequency of $0.034\,\mathrm{days^{-1}}$. 
	(Bottom) The $S$-index as a function of time. Note the long-term
	variations in the stellar activity.
	\label{fig:rotact}}
\end{figure*}

The EPD (pre-TFA) HATNet \lc\  of \hatcur\  shows significant periodic
variations with $P\approx29.2$~days and a peak-to-peak amplitude of 6.4
mmag in \band{I} (\reffig{rotact}). This period is detected both in the
auto-correlation function and the Fourier power spectrum of the light
curve. The period is detected in both seasons covered by the HATNet
light curve, and the signal remains in phase across both seasons as
well. While this period is suspiciously close to the lunar cycle, we
note that the \lc\ is relatively unchanged by the EPD procedure,
i.e.~these variations do not appear to correlate with any external
parameters, including the sky background. The TFA procedure does
suppress this signal when several hundred template stars are used.
However, this is often the case when applying non-reconstructive TFA to
other long-period variable stars such as Cepheids or Miras. To test
whether or not this signal is due to some non-astrophysical systematic
variation we searched for other light curves exhibiting a strong peak
in the Fourier power spectrum near $f = 0.034$\,[c/d]. We found that 9
out of the 5000 brightest stars with $I \la 10.0$ showed such a peak,
however none of these stars are in phase with \hatcur. Moreover, the
distribution of peak frequencies does not show a pile-up at $f = 1/P =
0.034$\,d$^{-1}$ relative to other frequencies, which one might expect
to see if this variation were a systematic trend. As a final test we
attempted to recover the signal after applying different TFA template
sets. We tried 100 disjoint template sets of 140 \lcs. In all cases the
$P\approx29.2$~day signal was recovered in the auto-correlation
function. We conclude that the signal is not a systematic variation
that is present in the light curves of many other stars, and if it is
not of astrophysical origin then it is due to a phenomenon that affects
the \lc\ of \hatcur\ in an apparently unique manner.

A likely interpretation of the variation seen in \reffig{rotact} is
that it is due to the rotational modulation of starspots on the surface
of \hatcur. The 6.4\,mmag amplitude of the variation is consistent with
other observations of spotted K dwarf stars. Note that there is no
significant $f_2 = 0.034$\,[c/d] periodicity in the RV data, when it is
modeled as the combination of a Keplerian orbit plus a sinusoid
component (see \reffig{rvfreq}). 
%
%
The secondary peaks in the autocorrelation function in the figure, plus
the co-phasing of the 29~day variation over two observing seasons (left
curve of \reffig{rotact}) indicate that individual starspots or
starspot groups persist for at least several rotations.

If the variation is indeed due to rotational modulation by starspots,
then the rotation period of \hatcur\ is $P\approx29.2$~days.  This may
be compared with the rotation period predicted from the $B-V$ color of
the star and the Ca II emission index $S$, using relations of
\cite{noyes:1984}. From \reftab{rvs}, the median value of $S$ observed
from HIRES is $\Savg = 0.61$. This value and $B-V = 1.046\pm0.024$ as
derived earlier yield a rotation period of $P_{calc} = 24.2$\,d. The
uncertainty in this calculation is difficult to quantify. The relations
given in \cite{noyes:1984} were based on a theoretically motivated fit
to the rotation period determined for a number of lower main sequence
stars from the rotational modulation of their chromospheric emission.
For the 18 K stars in their sample with measured rotation periods
$P_{obs}$, the rms difference between $P_{obs}$ and the calculated
value $P_{calc}$ is 3.5\,d. However, \citet{noyes:1984} pointed out
that the empirical relation for $P_{calc}$ as a function of $S$ and
$B-V$ becomes rather unreliable for $B-V\gtrsim1.0$, because of a
paucity of stars with observed rotation periods in this color range.
Hence the true uncertainty is doubtless somewhat larger than 3.5\,d. 
We conclude that a rotation period of 29.2\,d is consistent with
expectations from the star's $B-V$ color and \Savg\ value.

A 29~day rotation period of \hatcur, coupled with its radius, implies
an equatorial velocity $v_{eq} = 1.3\,\kms$.  Let us assume that the
stellar rotational axis is inclined at $i \sim 90^{\circ}$ to the line
of sight. This is supported by the geometry of our Solar system and the
generally similar geometries of the stellar systems with transiting
planets whose projected rotational axis inclinations have been measured
through the Rossiter-McLaughlin effect. We would then expect
$\vsini\sim1.3\,\kms$, consistent with the value of \vsini\ reported in
\reftab{stellar}.

\begin{figure}[!ht]
\plotone{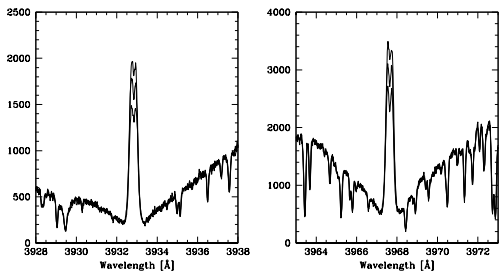}
\caption{
	The Calcium K (left) and H (right) line profile in selected HIRES
	observations of \hatcur. Both panels show three spectra overlaid;
	data taken at high, median and low activity, as characterized by
	the $S$ index. The spectra are matched to a common flux/wavelength
	scale using points outside the H and K line cores. The vertical
	axes on the plots are arbitrary and proportional to counts. 
	\label{fig:cahk}
}
\end{figure}

\reffig{rotact} shows the time history of the Ca II $S$ index. 
\reffig{cahk} shows the very prominent emission cores of the Ca II H
and K lines observed at three levels of activity during the time of the
HIRES observations.  A long-term variation of $S$ is apparent, with
timescale close to the length of the current data set.  As discussed
later, this may be due to long-term variations of stellar activity
analogous to the solar activity cycle. Phasing of the $S$ data at a
period of 29.2~days shows no evidence for periodic behavior. If the
star is really rotating at a period of 29~days, this suggests that the
chromospheric emission in the H and K lines is spread nearly uniformly
in longitude over the star.

The color $B-V=1.046$ of \hatcur\ and its median level of chromospheric
emission $\Savg = 0.61$ imply a chromospheric emission ratio $R'_{HK}$
given by log $R'_{HK} $ = -4.584 \citep{noyes:1984}.  From this we may
crudely estimate an age for the star using the inverse square root
relation between activity and age as originally posited by
\citet{skumanich:1972}. Using a fit of this relation to the Sun, the
Hyades, the Ursa Major group, and 412 individual lower main sequence
stars as derived by \citet{soderblom:1991}, we determine a
``chromospheric age'' $T_{cr}\sim$1.25\,Gyr.  The uncertainty in this
estimate is difficult to quantify, especially because most of the stars
in the sample had $B-V<1.0$, but it does suggest that based on its
chromospheric emission level the star is likely to be at the low end of
the age range given in the isochrone fit discussed in
\refsec{basicstel}.

\section{Excluding blend scenarios}
\label{sec:blend}

\subsection{Spectral line-bisector analysis}
\label{sec:bisec}

As always in determining whether the signature from combined
photometric and radial velocity variations in a star is due to a
transiting planet, it is necessary to exclude the possibility that the
entire combined information is due to a set of circumstances that
falsely give rise to the characteristic signature of a transiting
planet.  Following \cite{torres:2007}, we first explored the
possibility that the measured radial velocity variations are caused by
distortions in the spectral line profiles due to a nearby unresolved
faint eclipsing binary, whose relatively large radial velocity
variations mix with the non-varying spectrum of the primary and give
rise to an apparent radial velocity signal with the observed small
amplitude. In this case it has been shown \citep[e.g.][]{queloz:2001}
that the bisector span (BS) of spectral lines in the blended spectrum
varies in phase with the radial velocity signal itself, with similar
amplitude. We have carried out an analysis of the bisector span based
on the Keck spectra as described in earlier HATNet detection papers
\citep[see][]{bakos:2007a}, and do not see any statistically
significant correlation between the line bisector spans and the
measured radial velocities (confer \reffig{rvbis} and \reftab{rvs}),
thus providing no support for the hypothesis that the signal is caused
by a blended eclipsing stellar system.

However, because of the small amplitude of the RV signal (and hence
small amplitude of expected bisector span variations if the apparent RV
signal is indeed due to a blend), coupled with the large jitter as seen
in the RV residuals in \reffig{rvbis}, the above test cannot completely
rule out contamination by an unresolved binary system. Thus we sought
other, independent ways of excluding blend scenarios. In the following
subsections we attempt to model the system as a hierarchical triple,
where the smallest component is either a star or a Jupiter-sized
planet, and also as a background eclipsing binary blended with the
light of a foreground K dwarf. We show that none of these models are
consistent with all of the available observations.

\subsection{Detailed blend modeling of a hierarchical triple}
\label{sec:triple}

We consider the possibility that \hatcur\ is a hierarchical triple
system, having a deeper intrinsic eclipse of two bodies diluted by the
light of the bright K dwarf. Note that the bright K dwarf with well
known properties (parallax, proper motion, etc.) in this putative
triple system is referred to as \hatcur\ in the following discussion.
To rule out the hierarchical triple scenario, we attempt to fit a blend
model to the observations.  There are two scenarios for the eclipsing
system in the triple system. The first one is a stellar eclipsing
binary. The second case is a transiting hot Jupiter orbiting a low-mass
star, with the few percent deep transit diluted by \hatcur\ to become
only \hatcurLCdip\,mmag. We note that the shallower the transit event,
the larger the variety of configurations that can match the
observations. \hatcur\ represents a transition towards millimag transit
events, the blend analysis of which will certainly be challenging.

In each case (stellar binary, vs.~transiting hot Jupiter blend) we fit
the follow-up $z$, $r$ and $I_C$-band \lcs\, together with the HATNet
$I$-band \lc\ following a procedure similar to that described by
\citet{torres:2005}.  We assume that the bright star (\hatcur), which
is not eclipsed, has the mass, metallicity, age and distance determined
in \refsec{stelparam}. We also assume the components of the eclipsing
system have the same metallicity, age and distance as the bright star,
i.e.~they form a hierarchical triple. We fix the ephemeris, orbital
eccentricity and argument of periastron to the values determined in
\refsec{fitres} and vary the masses of the two components and the
inclination of the orbit. The downhill simplex algorithm is used to
optimize the free parameters. For the case where one of the components
is a planet, we vary the radius of the planet rather than its mass. For
the eclipsing binary case the magnitudes and radii of the stars are
taken from the Padova isochrones \citep{girardi:2000} applying the
radius correction described by \citet{torres:2005}. We use these
isochrones to allow for stars smaller than $0.4\,\msun$. For the planet
case we use the Yonsei-Yale isochrones to be consistent with the
single-star planet modeling.

\begin{figure}[!ht]
\plotone{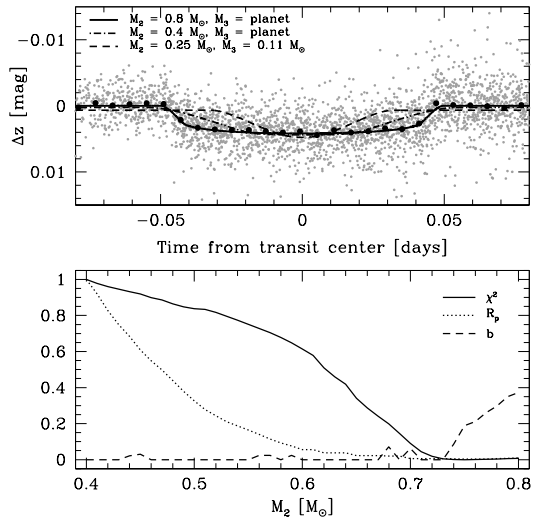}
\caption{
	(Top) The z-band follow-up \lc\  for \hatcur\ with model \lcs\ for
	three blend models over-plotted. The dark points show the binned
	\lc\ , while the light points show the unbinned light curve. The
	solid line shows a model where a planet orbits one component of a
	binary star, with the planet-hosting star having a mass of
	$0.8\,\msun$. The dot-dashed line shows a similar scenario but
	where the planet-hosting star has a mass of $0.4\,\msun$. The
	dashed line shows the best fit hierarchical triple model consisting
	of eclipsing binary star system bound to a distant third star. We
	can rule out the hierarchical triple scenario and the blend models
	where a planet orbits a binary star component that is smaller than
	$0.72\,\msun$, blend scenarios with a larger planet-hosting star
	can fit the \lc\ but a second star this massive would have been
	detected in the Keck and DS spectra.
	(Bottom) Here we show $\chi^{2}$, the planet radius $R_{p}$ and the
	impact parameter $b$ as a function of the planet-hosting star mass
	for the blend scenario of a planet orbiting one component of a
	binary system. $\chi^{2}$ is linearly scaled between 95788 and
	100457 while $R_{p}$ is linearly scaled between $0.583 R_{J}$ and
	$2.17 R_{J}$. For stellar masses below $0.72\,\msun$ the planet
	radius needed to fit the transit depth and duration yields an
	ingress and egress time that is too long.  Above $0.72\,\msun$ the
	impact parameter increases to fit the transit duration while the
	planet radius needed to fit the transit depth changes slowly. 
	\label{fig:blend}
}
\end{figure}

\reffig{blend} shows the best fit model for the eclipsing binary case
and two illustrative models for the blended hot-Jupiter case
over-plotted on the \band{z} \lc. We are unable to fit the \lc\ using a
combination of three, physically associated stars, but are able fit it
as a planet transiting one component of a binary star system, so long
as the star hosting the planet has $M > 0.72\,\msun$. Interestingly,
$0.72\,\msun$ emerges as a critical mass (and corresponding critical
radius, luminosity) for the eclipsed star, and splits the parameter
space into two disjoint domains.  As the mass of the host star is
reduced below $0.72\,\msun$ (and its radius is decreased), the only way
to maintain the depth and duration of the already central ($b=0$,
i.e.~longest possible) transit is to increase the radius of the planet.
This, however results in a longer ingress and egress time than is
allowed by the observations. Above $0.72\,\msun$ the radius of the
planet needed to fit the transit depth changes slowly, while raising
the impact parameter of the planet can compensate to fit the transit
duration (see \reffig{blend}). While a model of this form fits the \lc,
the $V$-band light ratio of the planet-hosting star to the brighter
$0.81\,\msun$ star would be $> 0.46$. The spectrum of a second star
this bright should have been easily detected in the Keck and DS
observations, and therefore we can rule out both of these blend
scenarios.

\subsection{Contamination from a background eclipsing binary}
\label{sec:bgeb}

\begin{figure}[!ht]
\plotone{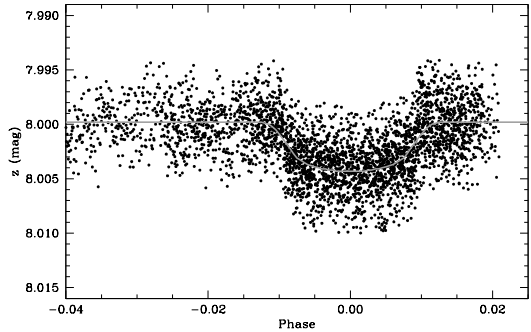}
\caption{
	Sample blend model fitted to our combined $z$-band measurements
	(arbitrary scale on the vertical axis). The solid line represents
	the \lc\ resulting from a background eclipsing binary system whose
	flux is diluted by the main star, which contributes 95\% of the
	$z$-band light in this case. The eclipsing binary is composed of
	0.64~$\msun$ and 0.13~$\msun$ stars in an edge-on configuration,
	placed 60~pc behind the target. The predicted relative brightness
	of the primary in this scenario is only 3\% of the light of the
	main star in the optical.\label{fig:bgeb}
}
\end{figure}

An alternate model of an astrophysical false positive that would make
the contaminating star much fainter relative to the main object is one
involving a background eclipsing binary (chance alignment). We modeled
this case using the same methodology described above, fitting the
combined $z$-band follow-up \lcs\ as the sum of the light from three
stars. The properties of the main star were held fixed as before, and
those of the eclipsing binary components were constrained to lie on the
same isochrone, which for simplicity was taken to be the same as the
main star. Extensive tests indicated that the detailed shape of the
resulting synthetic \lc\ is fairly insensitive to the distance at which
we place the eclipsing binary behind the target, and thus acceptably
good fits to our $z$-band photometry can be achieved for a wide range
of distances, given the measurement precision. \reffig{bgeb} shows an
example of such a fit, in which the eclipsing binary is placed about
60\,pc behind the main star and is composed of a 0.64\,$\msun$ star
(spectral type K7V--M0V) orbited by a 0.13\,$\msun$ stellar companion.
This model is not only consistent with the photometry, but it predicts
an optical brightness for the primary of the eclipsing binary of only
3\% relative to the main star. Detecting such a faint set of spectral
lines in our Keck spectra would be challenging. Furthermore, if we
place the eclipsing pair 110\,pc behind the main star, the \lc\ fit is
still about the same, but the relative brightness decreases by a factor
of two, making the eclipsing binary spectroscopically undetectable.
Spectral line bisector variations predicted by this second model
\citep[see, e.g.,][]{torres:2005} would be at the level of the scatter
in our actual measurements, and thus could not be entirely ruled out
either.

\begin{figure}[!ht]
\plotone{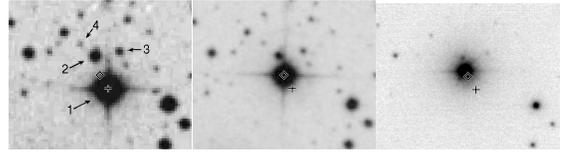}
\caption{
	Images of a $2\arcmin\times1.7\arcmin$ field containing \hatcur\
	from the POSS-I Red survey (Left), POSS-II Red survey (Center) and
	a \flwof\ $z$-band image (Right, see \refsec{phfu}). The dates of
	the exposures are 10 July 1951, 25 August 1989, and 2 September
	2007 respectively. The cross marks the position of \hatcur\ in the
	POSS-I image (labeled as star 1), while the diamond marks the
	position of \hatcur\ in the POSS-II image.  Between 1951 and 2007
	HTR155-001 moved 13.8\arcsec\ in the NNE direction.  Stars labeled
	1 through 4 on the POSS-I image have USNO A-2.0 R magnitudes of
	10.8, 15.1, 17.2 and 18.9 respectively
	\citep[USNO-A2.0;][]{monet:1998}. From the POSS-I image we can rule
	out stars brighter than $R\sim19$ at the current position of
	\hatcur.
\label{fig:poss}}
\end{figure}

%
While this blend scenario appears to satisfy all observations, it
implies the presence of an eclipsing binary $\sim$4.5 magnitudes
fainter in the optical very near the present location of our target. It
is fortunate that \hatcur\ has a large proper
motion\citep[$0\farcs264$~yr$^{-1}$;][]{perryman:1997}. Using Palomar
Observatory Sky Survey plates from 1951 (POSS-I, red and blue plates),
we can view the sky at the current position of \hatcur\ {\em
unobstructed}, since the target was 15\arcsec\ away. \reffig{poss}
shows a stamp from the POSS-I, POSS-II plates, and also a current
observation with the \flwof\ telescope. The proper motion of \hatcur\
is apparent. A number of faint stars are marked on the figure. Note
that the faintest possible star that can cause the observed
\hatcurLCdip\ mmag variation in the combined \lc\ would be 6 magnitudes
fainter than \hatcur, and in this extreme situation the blended faint
star would need to completely disappear during its transit. There is no
faint star down to $\sim 19$\,mag within $\sim 5\arcsec$ of the current
position of \hatcur. The closest star is marked ``2'' in \reffig{poss}
is 2MASS~19505049+4805017, with $z\approx 14.4$, or about $z\approx5.6$
magnitudes fainter than \hatcur. This is well resolved in some of our
follow-up photometry observations, such as on the 2007 September 2
night, and its brightness was constant, with r.m.s.~much smaller than
the required amplitude of its variation should be ($\sim 1.3$\,mag) to
cause the observed \hatcurLCdip\ millimag combined dip.

In summary, i) we have not seen bisector variations correlated with the
radial velocity, ii) we investigated unresolvable hierarchical triple
systems (blended eclipsing binary or transiting hot Jupiter), and found
all to be incompatible with the photometric and spectroscopic data
together, and iii) we were able to exclude chance alignment of a
background eclipsing binary based on the high proper motion of the
star. We therefore conclude that the transit signal, and the synchronized
RV signal, indeed are both due to a sub-stellar companion transiting
the \hatcurYYspec\ dwarf \hatcur.


\section{Global modeling of the data}
\label{sec:globmod}

In order to perform an optimal analysis of the data at hand, we
assembled a global model of a planetary transit scenario that describes
all the data components: i) the HATNet discovery \lc, ii) the
photometric follow-up \lcs\ of 10 independent events, and iii) the
radial velocity data from Keck (\refsec{modphys}). The physical model
was then extended by a model that describes systematic variations in
the data (\refsec{modsys}). We then performed a joint fit of the
combined physical plus systematic model by using all these data
simultaneously to derive the physical parameters, and also to correct
for the systematic variations (\refsec{fitproc}). The results are
discussed in \refsec{fitres} and \refsec{kh}.

\subsection{Physical model for the \lc\  and radial velocity}
\label{sec:modphys}

Our model for the follow-up \lcs\ used analytic formulae based on
\citet{mandel:2002} for the eclipse of a star by a planet, where the
stellar flux is described by quadratic limb-darkening. We did not
attempt to fit for the limb darkening coefficients, as they require
very high quality data to resolve their degeneracy with other
parameters. Instead, the limb darkening coefficients were derived from
the second iteration SME results (\refsec{stelparam}), using the tables
provided by \citet{claret:2004} for $z$, $r$ (\flwof) and $I_C$ bands
(\piszkessch). The transit shape was parametrized by the normalized
planetary radius $p\equiv \rpl/\rstar$, the square of the impact
parameter $b^2$, and the reciprocal of the half duration of the
transit\footnote{
	The duration is defined as the time between the contact centers,
	i.e.~when the center of the planetary disk crosses the limb of the
	star during ingress and egress.
}
$\zrstar$. We chose these parameters because of their simple geometric
meanings and the fact that these show negligible correlations
\citep[see][]{carter:2008,bakos:2007b,pal:2008b}.

Our model for the HATNet data was a simplified version of the
\citet{mandel:2002} analytic functions, because i) the number of
in-transit data points in the HATNet \lc\ is significantly less than
the same number for the follow-up \lcs, and ii) the individual errors
on the points is much worse for HATNet than for the follow-up. Thus,
while the HATNet data are efficient in constraining the ephemeris, they
are not comparable to the follow-up data in determining other transit
parameters that depend on the exact shape of the transit (such as
depth, or duration of ingress).  For these reasons, we employed a
simple transit model that neglects limb darkening. This model function,
which we label as ``P1P3'', originates from an expansion of the exact
\citet{mandel:2002} model function by Legendre polynomials, resulting
in a functional form (omitting scaling factors) of the ingress (and
symmetrically, the egress) of $x\to (21x-5x^3)/16$, where the
independent variable $x$ is scaled to be $1$ at first (third) contact,
and $-1$ at second (fourth) contact. For reference, the trapeze model
has a simple linear function describing the ingress: $x\to x$. The
``P1P3'' model approximates the \citet{mandel:2002} function to better
than 1\%, and our experience shows that such an approximation yields
the same timing precision as more sophisticated model functions. In
addition, using the simplified model yields much faster computation. A
similar transit model approximation using hyperbolic tangent functions
was described earlier by \cite{protopapas:2005}. The depth of the
HATNet transits was also adjusted in the fit (independent of the
follow-up data) and we found that the depth derived from the HATNet
\lc{} is comparable (\refsec{fitres}) to the depth of the (EPD and TFA
reconstructed) follow-up transit measurements, namely the ratio of the
two was $B_{\rm inst}=0.95\pm0.08$.  In general, this ``instrumental
blend'' parameter has to be adjusted independently since the possible
contamination by nearby stars in the wide-field images may yield
shallower transits.

The adopted model for the radial velocity (RV) variations was described
earlier in \refsec{hires}. The RV curve was parametrized by an
eccentric Keplerian orbit with semi-amplitude $K$, RV zero-point
$\gamma$, Lagrangian orbital elements
$(k,h)=e\times(\cos\omega,\sin\omega)$, plus a linear trend $G_1$.

We assumed that there is a strict periodicity in the individual transit
times. In practice, we introduced the first transit center as $T_{\rm
A}=T_{c,-231}$ and the last transit center as $T_{\rm B}=T_{c,+91}$,
covering all of our measurements with the HATNet telescopes, the
\flwof\ telescope and the Schmidt telescope. The transit center times
for the intermediate transits were interpolated using these two epochs
and the $N_{tr}$ transit number of the actual event under the
assumption of no transit timing variations. The model for the RV data
contained the ephemeris information through the $T_{c,-231}$ and
$T_{c,+91}$ variables; for instance the period assumed during the fit
was $P = (T_{c,+91} - T_{c,-231})/ (91 + 231)$.
The other coupling between the transit
photometry and the RV data is via the $k$ and $h$ Lagrangian orbital
elements that determine the relation between the photometric and
orbital ephemeris, plus have a minor effect on the transit shape.
Altogether, the 12 parameters describing the physical model were
$T_{c,-231}$, $T_{c,+91}$, $\rpl/\rstar$, $b^2$, $\zrstar$, $K$,
$\gamma$, $k = e\cos\omega$, $h = e\sin\omega$ and $G_1$.  As mentioned
above, the parameters were extended with the instrumental blend factor
$B_{\rm inst}$, and the HATNet out-of-transit magnitude, $M_{\rm
0,HATNet}$.

\subsection{Models for systematic variations}
\label{sec:modsys}

A ``joint'' fit of the physical model to the global data-set has been
performed routinely for recent discoveries
\citep{bakos:2008,latham:2008}. Here we extended our physical model
that describes the transit events and the radial velocities with an
instrumental model that describes the systematic variations of the
data. Since the HATNet photometry has been already EPD and TFA
corrected (\refsec{disco}), and the Keck RVs have been investigated for
trends (\refsec{hires}), but deemed to have insufficient number to
decipher any significant systematic variation (50 points compared to
$\ordo(10^4)$ points for photometry), we only modeled systematic
variations of the follow-up photometry (\refsec{phfu}). As mentioned
earlier, the effect of systematics on the shallow transit in the
photometric follow-up data is enhanced, and can impact the basic
planetary parameters, such as the planetary radius. We experimented
with several models, since the treatment of systematics may improve
precision, but not necessarily the accuracy of the results. For
reference, we call the simplest model, with no EPD and no TFA, as model
``{\tt E0T0}'' (for technical details, see \refsec{sysvar}).

Model ``{\tt E0TRG}'' was independent (i.e.~not simultaneous)
reconstructive global TFA of the \flwof\ \band{z} data (8 nights, see
\reftab{phfu}), assuming a trapeze shaped $m_0(\vec{p},t)$ model
function in \refeq{tfamin} with three parameters (total duration,
ingress duration, depth). The period was iteratively refined using both
the \flwof\ and the HATNet data, but reconstructive TFA was only
performed on the FLWO photometry. We used the magnitude values
transformed to the selected reference frame for each night (see
\refsec{phfu}), and the zero-point shift for each night was determined
separately. We increased the number of template stars till the S/N of
the trapeze reached its maximum (S/N = 39), requiring 246 stars with
$z\lesssim14$. Note that this was a global TFA, in the sense that there
was one $c_k$ TFA coefficient per star in \refeq{tfafilter} for all
nights combined (see \refsec{tfa}). The reconstructed \lc\ was then fed
into the joint fit.

Model ``{\tt ELT0}'' had EPD parameters in the simultaneous fit,
together with the parameters of the physical model, but no TFA-based
detrending, i.e.~all the systematic variations in the follow-up
photometry were modeled as due to external parameters. We chose 5 such
parameters, namely the hour angle (characterizing a monotonic trend
that linearly changes over time), the square of the hour angle, the
stellar profile sharpness parameter, $S=(2.35/{\rm FWHM})^2$, the X and
Y pixel position on the chip (each with two coefficients). The exact
functional form of the above parameters contained 8 coefficients,
including the auxiliary out-of-transit magnitude of the individual
events. The EPD parameters were independent for all 10 nights, implying
80 additional coefficients in the global fit.

Model ``{\tt E0TL}'' had local TFA parameters in the instrumental model
for each night, and no EPD was used. We selected 23 TFA template stars
that were present in all observations, representing faint and bright
stars spread across the chip. The total number of $c_k$ coefficients in
the simultaneous fit was thus 230.
Model ``{\tt E0TG}'' had global TFA parameters for all nights, using
211 stars that were present on all frames. Note that without
simultaneous fit using an underlying physical model ($m_j(t_i))$ in
\refeq{tfamin}) that takes into account the different shape of the
\lcs{} acquired through different photometric bands, such a global TFA
fit would be not possible.

We tested variants of the above methods. Model ``{\tt ELTL}'' was a
simultaneous fit of EPD and TFA parameters in local EPD-mode and local
TFA-mode with altogether $80 + 230$ parameters in addition to the 12
physical parameters. Model ``{\tt ELTG}'' was a simultaneous fit of
local EPD and global TFA parameters with an additional $80 + 211$
instrumental parameters.  The parameter sets were extended in all cases
with the out-of-transit magnitudes for each follow-up \lc\ and the
HATNet photometry. The number of fitted parameters was much smaller
than the number of data points ($\sim 5000$).

\subsection{Performing the joint fit}
\label{sec:fitproc}

It is computationally rather intensive to perform a fit with $\sim 300$
parameters and to determine their respective error distributions.

Since the majority of the fitted parameters appear as linear terms in
the final form of the model functions, we minimized \chisq\ in the
parameter space by using a hybrid algorithm, namely by combining the
downhill simplex method \citep[a.k.a.~AMOEBA, see][]{press:1992} with
the classical linear least squares algorithm. The simplex itself was
propagating in the ``nonlinear'' hyperplane of the parameter space,
while in each point of the hyperplane the value of $\chisq$ was
computed using classical linear least squares (CLLS) minimization. This
method yielded a very good convergence even with the large number of
free parameters.

Uncertainties of the parameters were derived using the Markov Chain
Monte-Carlo method \citep[MCMC, see][]{ford:2006}, by starting two
types of chains from the best fit value.  The first one was the
classical MCMC chain, letting all parameters vary. The other type of
chain was generated by only allowing the variation in a hyperplane of
the parameter space consisting of the nonlinear parameters and of some
additional linear parameters where error-calculation was desired
(examples are $K$, $\gamma$ or $G_1$). In this ``Hyperplane-CLLS''
chain the remaining linear parameters were determined at each step via
CLLS.  The H-CLLS chain has the advantage that transition probabilities
are much higher in the separated hyperplane, and thus the convergence
and computing time requirements are also smaller by an order of
magnitude. Furthermore, the error distribution of parameters for the
hyperplane variables was identical to those determined by classical
MCMC. The distribution for the parameters outside the hyperplane was
distorted, but these were considered as auxiliary, where such
distribution is irrelevant (e.g.~the EPD and TFA coefficients).
Therefore the final distribution of the physical parameters was derived
by using H-CLLS chains.

For both types of the chains the \emph{a priori} distributions of the
parameters were chosen from a generic Gaussian distribution, with
eigenvalues and eigenvectors derived from the Fisher covariance matrix
for the best fit value. All functions that appear in the physical and
instrumental model are analytical, and their partial derivatives are
known. The parametric derivatives of the transit \lc{} model of
\cite{mandel:2002} are found in \cite{pal:2008b}, while the parametric
derivatives of the eccentric radial velocity curves can be obtained by
involving the implicit function theorem, and can be expressed as
functions of the eccentric anomaly.
%
%
Thanks to these analytic properties, the derivation of the Fisher
matrix was straightforward.

\subsection{Results of the fit and the effect of systematics}
\label{sec:fitres}

\ifthenelse{\boolean{emulateapj}}{
	\begin{deluxetable*}{lrrrr}
}{
	\begin{deluxetable}{lrrrr}
}
\tablewidth{0pc}
\tablecaption{\lc\  parameters involving different 
	kinds of models for systematic variations.
	\label{tab:trendparam}
}
\tablehead{
	\colhead{Method} &
	\colhead{\ensuremath{T_{c,+91}-2454790.0}}	&
	\colhead{\ensuremath{R_{\rm p}/R_\star}} &
	\colhead{\ensuremath{b^2}}		&
	\colhead{\ensuremath{\zeta/R_\star}}
}
\startdata
	{\tt E0T0}\tablenotemark{a}  & $1.62781\pm0.00055$   & $0.06035\pm0.00140$ & $0.165\pm0.120$ & $22.05\pm0.24$    \\
	{\tt E0TL}\tablenotemark{b}  & $1.62793\pm0.00047$   & $0.05831\pm0.00094$ & $0.114\pm0.087$ & $21.82\pm0.18$    \\
	{\tt E0TG}\tablenotemark{c}  & $1.62772\pm0.00046$   & $0.05730\pm0.00096$ & $0.112\pm0.091$ & $21.99\pm0.19$    \\
    {\tt E0TRG}\tablenotemark{d} & $1.62758\pm0.00133$   & $0.05690\pm0.00150$ & $0.223\pm0.137$ & $22.10\pm0.27$    \\
	{\tt ELT0}\tablenotemark{e}  & $1.62834\pm0.00044$   & $0.05825\pm0.00094$ & $0.109\pm0.093$ & $22.25\pm0.17$    \\
	{\tt ELTL}\tablenotemark{f}  & $1.62826\pm0.00040$   & $0.05675\pm0.00097$ & $0.113\pm0.093$ & $22.17\pm0.16$    \\
	{\tt ELTG}\tablenotemark{g}  & $1.62832\pm0.00039$   & $0.05758\pm0.00091$ & $0.120\pm0.087$ & $22.15\pm0.15$    \\
\enddata
\tablenotetext{a}{No EPD and no TFA performed on the photometric
                  follow-up data.}
\tablenotetext{b}{No EPD, and per-night ``local'' TFA performed with 23 TFA 
                  template stars and simultaneous fitting with the transit 
                  model.}
\tablenotetext{c}{No EPD, and global simultaneous TFA performed with 
                  211 TFA templates.}
\tablenotetext{d}{No EPD. Global reconstructive trapeze TFA 
                  performed with 246 TFA templates. Only the 8 nights
                  of FLWO \band{z} observations were used.}
\tablenotetext{e}{EPD performed using 5 free parameters per night. No TFA.}
\tablenotetext{f}{EPD as for {\tt ELT0}. TFA as for {\tt E0TL}.}
\tablenotetext{g}{EPD as for {\tt ELT0}. TFA as for {\tt E0TG}.}
\ifthenelse{\boolean{emulateapj}}{
	\end{deluxetable*}
}{
	\end{deluxetable}
}

We performed a joint fit on the HATNet \lc, the follow-up photometry,
and the Keck radial velocities (the ``data''), using the physical model
described in \refsec{modphys}, as extended by the various models for
systematics (\refsec{modsys}), and using the algorithms described in
\refsec{fitproc}.

We carried out the fit considering each model describing the
systematics in the follow-up photometry separately in order to
investigate their effect on the final results. As mentioned earlier,
this is important due to the very shallow transit signal and the large
relative amplitude of systematic variations. We used the $m_f$ fitted
magnitude values (\refsec{disco}) for the photometry follow-up \lcs\
(without prior EPD or TFA). Since only the follow-up photometric
data-component was described by the different systematic models (see
\refsec{modsys}), we focused on those \lc\ parameters that are affected
by the photometric follow-up, namely the transit center time of the
last event $T_{c,+91}$, the radius ratio of the planet to star, the
(square of the) impact parameter and the transit duration parameter
$\zrstar$.  The other adjusted parameters were not relevant in this
comparison since they are not affected by the follow-up \lcs. Note,
that with the exception of the {\tt E0TRG} model, all EPD and TFA
parameters were fit {\em simultaneously} with the physical model
parameters.

%
The best fit values and the respective uncertainties are summarized in
\reftab{trendparam}. It is clear that EPD and TFA both significantly
reduce the uncertainties in all the parameters. Note that the
implementation of {\tt E0TRG} was somewhat sub-optimal. Because the
reconstructive trapeze shape was not bandpass-dependent (as opposed to
the full \citet{mandel:2002} model used in the other cases), only a
fraction of the follow-up was used, namely all photometry in \band{z}.
When using EPD and TFA together, the global TFA ({\tt ELTG}) performs
better than local TFA ({\tt ELTL}) for each night. The improvements by
EPD ({\tt ELT0}) and TFA ({\tt E0TG}) separately are roughly the same.
The best model, based on the formal error-bars, is the EPD and global
TFA together ({\tt ELTG}); here the unbiased error in the \lc\
parameters decreased by a factor of 1.5.

The final value for the parameters also changed by 3-$\sigma$ in the
sense that both $\rpl/\rstar$ and $b^2$ decreased. This means that the
treatment of the systematics recovers a sharper transit with smaller
impact parameter and smaller planetary radius. If the stellar
parameters were determined using the \arstar\ luminosity constraint
(\refsec{arstar}), then ignoring the correction for systematics would
lead to higher impact parameter, larger $\rpl/\rstar$, and smaller
\arstar, corresponding to smaller density (\refeq{arstar}). One may
expect that the effect of systematics is slowly diminished by
accumulating more data taken with different instruments. This is
consistent with our experience, that is the initial follow-up \lcs\
without proper treatment of systematics indicated larger impact
parameters, and a lower density dwarf star that was inconsistent with
the Hipparcos-based stellar parameters (\refsec{basicstel}). By
accumulating more data this inconsistency diminished (\refsec{arstar}).
Because both EPD and TFA point toward this direction, and also deliver
the smallest formal uncertainty, we expect that the most {\em accurate}
model will be the {\tt ELTG} in \reftab{trendparam}. This conjecture
will be decided with accurate ground-based follow-up photometry
\citep[such as presented recently in ][]{johnson:2008b}, and ultimately
when the {\em Kepler} space mission returns the \lc\ for \hatcur. This
will have important implications on optimal trend removal for other
transiting systems, where such a space-born ``reference'' will not be
available.

\begin{figure}[!ht]
\plotone{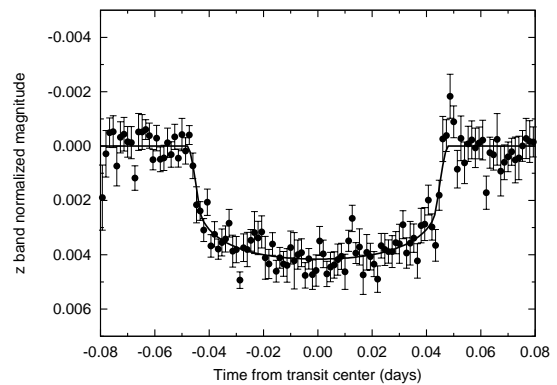}
\caption{
	Combined and binned \lc\  of \hatcur{}, involving
	all the $z$ band photometry, obtained by KeplerCam.
	The \lc{} is superimposed with the best fit model
	(which is the result the joint fit described in \refsec{globmod}).
\label{fig:fubin}}
\end{figure}

The results for the simultaneous fit using the {\tt ELTG} systematic
model for the follow-up \lcs\ were the following:
$T_{\mathrm{c},-231}=2453217.75466\pm0.00187$~(BJD), 
$T_{\mathrm{c},+91}=\hatcurLCT$~(BJD), 
$K=\hatcurRVK$\,\ms,
$k\equiv e\cos\omega=\hatcurRVk$,
$h\equiv e\sin\omega=\hatcurRVh$,
$\rpl/\rstar=\hatcurLCrprstar$, 
$b^2=\hatcurLCbb$,
$\zrstar=\hatcurLCzeta$\,$\mathrm{day^{-1}}$, and
$\gamma=\hatcurRVgamma$\,\ms,
$B_{instr}=0.95\pm0.08$,
$G_1 = \hatcurRVlindrift$\,\ms/day,
$M_{\rm 0,HATNet}=8.35892\pm0.00003$ (instrumental HATNet 
out-of-transit magnitude).
The combined and phase-binned \lc\ of all \band{z} \flwof\ observations
is shown on \reffig{fubin}. The planetary parameters and their
uncertainties can be derived by the direct combination of the \emph{a
posteriori} distributions of the \lc, radial velocity and stellar
parameters \citep[see also][]{pal:2008a}. We found that the mass of the
planet is $\mpl=\hatcurPPmlong\,\mjup = \hatcurPPmelong\,\mearth$, the
radius is $\rpl=\hatcurPPrlong\,\rjup = \hatcurPPrelong\,\rearth$ and
its density is $\rho_p=\hatcurPPrho$\,\gcmc. The final planetary
parameters are summarized at the bottom of Table~\ref{tab:parameters}.

\begin{deluxetable}{lc}
\tablewidth{0pc}
\tablecaption{Orbital and planetary parameters\label{tab:parameters}}
\tablehead{\colhead{~~~~~~~~~~~~~~~Parameter~~~~~~~~~~~~~~~} & \colhead{Value}}
\startdata
\sidehead{\Lc{} parameters}
~~~$P$ (days)             \dotfill    & $\hatcurLCP$              \\
~~~$T_c$ (${\rm BJD}$)    \dotfill    & $\hatcurLCT$              \\
~~~$T_{14}$ (days)
      \tablenotemark{a}   \dotfill    & $\hatcurLCdur$            \\
~~~$T_{12} = T_{34}$ (days)
    \tablenotemark{a}     \dotfill    & $\hatcurLCingdur$         \\
~~~$\arstar$              \dotfill    & $\hatcurPPar$             \\
~~~$\zrstar$              \dotfill    & $\hatcurLCzeta$           \\
~~~$\rpl/\rstar$          \dotfill    & $\hatcurLCrprstar$        \\
~~~$b \equiv a \cos i/\rstar$
                          \dotfill    & $\hatcurLCimp$            \\
~~~$i$ (deg)              \dotfill    & $\hatcurPPi$ \phn         \\

\sidehead{Spectroscopic parameters}
~~~$K$ (\ms)              \dotfill    & $\hatcurRVK$              \\
~~~$\gamma$ (\kms)        \dotfill    & $\hatcurRVgamma$          \\
~~~$G_1$ (\ms/day)           \dotfill    & $\hatcurRVlindrift$       \\
~~~$k_{\rm RV}$\tablenotemark{b} 
                          \dotfill    & $\hatcurRVk$              \\
~~~$h_{\rm RV}$\tablenotemark{b}
                          \dotfill    & $\hatcurRVh$              \\
~~~$k_C$\tablenotemark{c} \dotfill    & $\hatcurRVkcorr$          \\
~~~$h_C$\tablenotemark{c} \dotfill    & $\hatcurRVhcorr$          \\
~~~$e$                    \dotfill    & $\hatcurRVeccen$          \\
~~~$\omega$               \dotfill    & $\hatcurRVomega^\circ$    \\

\sidehead{Secondary eclipse parameters\tablenotemark{d}}
~~~$T_s$ (BJD)            \dotfill    & $\hatcurXsecondary$       \\
~~~$T_{s,14}$             \dotfill    & $\hatcurXsecdur$          \\
~~~$T_{s,12}$             \dotfill    & $\hatcurXsecingdur$       \\

\sidehead{Planetary parameters}
~~~$\mpl$ ($\mjup$)       \dotfill    & $\hatcurPPmlong$    \\
~~~$\rpl$ ($\rjup$)       \dotfill    & $\hatcurPPrlong$    \\
~~~$C(\mpl,\rpl)$
    \tablenotemark{d}     \dotfill    & $\hatcurPPmrcorr$   \\
~~~$\rhopl$ (\gcmc)       \dotfill    & $\hatcurPPrho$      \\
~~~$a$ (AU)               \dotfill    & $\hatcurPParel$     \\
~~~$\log g_p$ (cgs)       \dotfill    & $\hatcurPPlogg$     \\
~~~$T_{\rm eq}$ (K)       \dotfill    & $\hatcurPPteff$     \\
~~~$\Theta$               \dotfill    & $\hatcurPPtheta$    \\
~~~$F_{per}$ (\ergscmsq) \tablenotemark{e}
                          \dotfill    & $\hatcurPPfluxperi$ \\
~~~$F_{ap}$  (\ergscmsq) \tablenotemark{e} 
                          \dotfill    & $\hatcurPPfluxap$   \\
~~~$\langle F \rangle$ (\ergscmsq) \tablenotemark{e}
                          \dotfill    & $\hatcurPPfluxavg$  \\
\enddata
\tablenotetext{a}{%
	\ensuremath{T_{14}}: total transit duration, time
	between first to last contact; 
	\ensuremath{T_{12}=T_{34}}:
	ingress/egress time, time between first and second, or third and fourth
	contact.}
\tablenotetext{b}{Lagrangian orbital elements, based purely on the RV data.}
\tablenotetext{c}{Refined value of $h$ and $k$, 
	derived from RV analysis and the constraint given by the 
	$C_{\rm kh}\equiv(\sqrt{1-e^2}/(1+h)$ value, resulting from the \lc\  
	modeling and stellar evolution analysis.
}
\tablenotetext{d}{
	Correlation coefficient between the planetary mass \mpl\ and radius
	\rpl.
}
\tablenotetext{e}{
	Occultation parameters and incoming flux per unit surface area in
	periastron, apastron, and average for the orbit were calculated
	using the refined values $k_C$ and $h_C$.
}
\end{deluxetable}

\subsection{Constraints on orbital eccentricity}
\label{sec:kh}

\begin{figure}[!ht]
\plotone{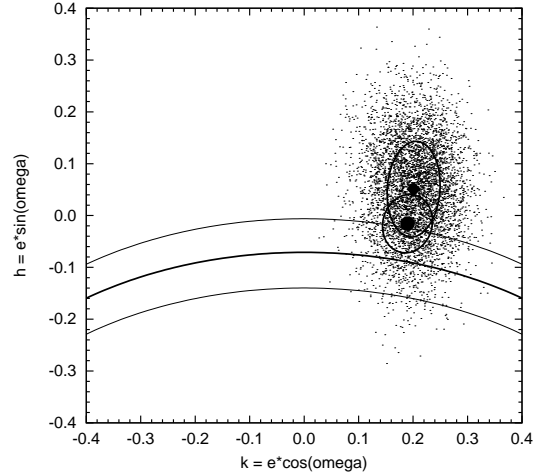}
\caption{
	The cloud of points represent the \emph{a posteriori} distribution
	of the orbital elements $k=e\cos\omega$ and $h=e\sin\omega$ yielded
	by the global modeling (\refsec{fitres}).  The best fit $(k,h)$
	pair is marked as a filled diamond, encircled with the 1-$\sigma$
	confidence ellipsoid. The constraint given by the
	$C_{kh}\equiv\sqrt{1-e^2}/(1+h)$ quantity defines a stripe in the
	$(h,k)$ plane. The location of the finally accepted $(k,h)$ value
	and its 1-$\sigma$ confidence are denoted by the dot below the
	diamond, and the surrounding ellipse, respectively.
\label{fig:kh}}
\end{figure}

An interesting aspect of the \hatcur\ system is that the \lc\ analysis,
the Hipparcos parallax and the theoretical isochrones together provide
an extra constraint on the orbital eccentricity. In our analysis the
orbital eccentricity and argument of pericenter are characterized by
the Lagrangian orbital elements $k=e\cos\omega=\hatcurRVk$ and
$h=e\sin\omega=\hatcurRVh$. If the analysis relied purely on the radial
velocity data, then the errors in $k$ and $h$ would be similar, the
error of $h$ being somewhat smaller (see \refsec{hires}).  However,
when the data are complemented with photometry of transit events (as in
our case), the period and phase of the orbit become tightly
constrained. As a result, the $k$ and $h$ orbital elements will have
different uncertainties since the mean longitude at the transit is
constrained more by $\cos\omega$ than $\sin\omega$ (this is due to the
same underlying reason that the phase lag between the secondary and
primary transits is also proportional to $e\cos\omega$ and does not
strongly depend on $e\sin\omega$). Indeed, the {\em a posteriori}
distribution of $h$ and $k$ as yielded by the joint analysis
(\refsec{fitres}) is asymmetric, and the uncertainty in $k$ is half of
the uncertainty in $h$ (\reffig{kh}, large oval around the filled
diamond).

By re-arranging \refeq{zetavsar}, we can define $C_{kh}$ as:
\begin{equation}
	\label{eq:kh}
	C_{\rm kh}\equiv\frac{\sqrt{1-h^2-k^2}}{1+h}=
	\frac{a}{\rstar}\left(\frac{\zeta}{\rstar}\right)^{-1}\frac{n}{\sqrt{1-b^2}}, 
\end{equation}
where $n=2\pi/P$ is the mean motion \citep[see also][]{ford:2008}. 
The period $P$, impact parameter
$b$, and \zrstar\ are well constrained by the photometry of multiple
transit events. The normalized semi-major axis \arstar\ is determined
from the apparent magnitudes, Hipparcos parallax and isochrones, as
described in \refsec{arstar}. These together yield an estimate of
$C_{kh} = \hatcurRVckh$, independent of the RV data, and thus provide
an additional constraint on $h$ and $k$ in the form of a hyperbola on
the $k-h$ plane. Taking into account the uncertainties in $C_{kh}$,
this appears as a stripe in the $k-h$ plane (\reffig{kh}). This
constraint can be used to decrease the uncertainty of $h$ by a factor
of two, so that it becomes similar to that of $k$.


In practice, we have a Monte Carlo distribution for $k$, $h$ and
$C_{kh}$. For each $k$ and $h$ value in MC distribution, we determine
the closest $k',h'$ point of the hyperbola (there is a unique solution
for this). Since the value for $C_{kh}$ constrains only one degree of
freedom, we can treat $(k,h)$ and $(k',h')$ as independent. Therefore,
we define $k_{\rm C}$ and $h_{\rm C}$ as the mean of $k$, $k'$ and $h$,
$h'$, respectively, and consider these as an improved estimate of the
orbital eccentricity parameters. The result is an MC distribution for
$k_{\rm C}$ and $h_{\rm C}$, where their mean values and errors can be
derived: $k_{\rm C}=\hatcurRVkcorr$ and $h_{\rm C}=\hatcurRVhcorr$.
Indeed, the errors in $k_{\rm C}$ and $h_{\rm C}$ are now comparable.

Based on $k_C$ and $h_C$, the refined orbital eccentricity is
$e=\hatcurRVeccen$ and the argument of pericenter is
$\omega=\hatcurRVomega^\circ$. Note that the above method is unique,
and can be applied for eccentric transiting systems with well-known
parallax but poorly determined RV orbit. It is likely that the {\em
Kepler} mission will find similar cases. A counter-example is HAT-P-2b
\citep{bakos:2007b}, where the RV amplitude is large compared to the
uncertainties (1\,\kms), and the error in the parallax is larger, thus
$h$ can not be refined. The current analysis points towards an even
more general way of simultaneously fitting to {\em all} data, including the
stellar isochrone search with various constraints (e.g.~parallax
information with errorbars).

\section{Discussion}
\label{sec:disc}

\subsection{The Planet \hatcurb}

\begin{figure*}[!ht]
\plotone{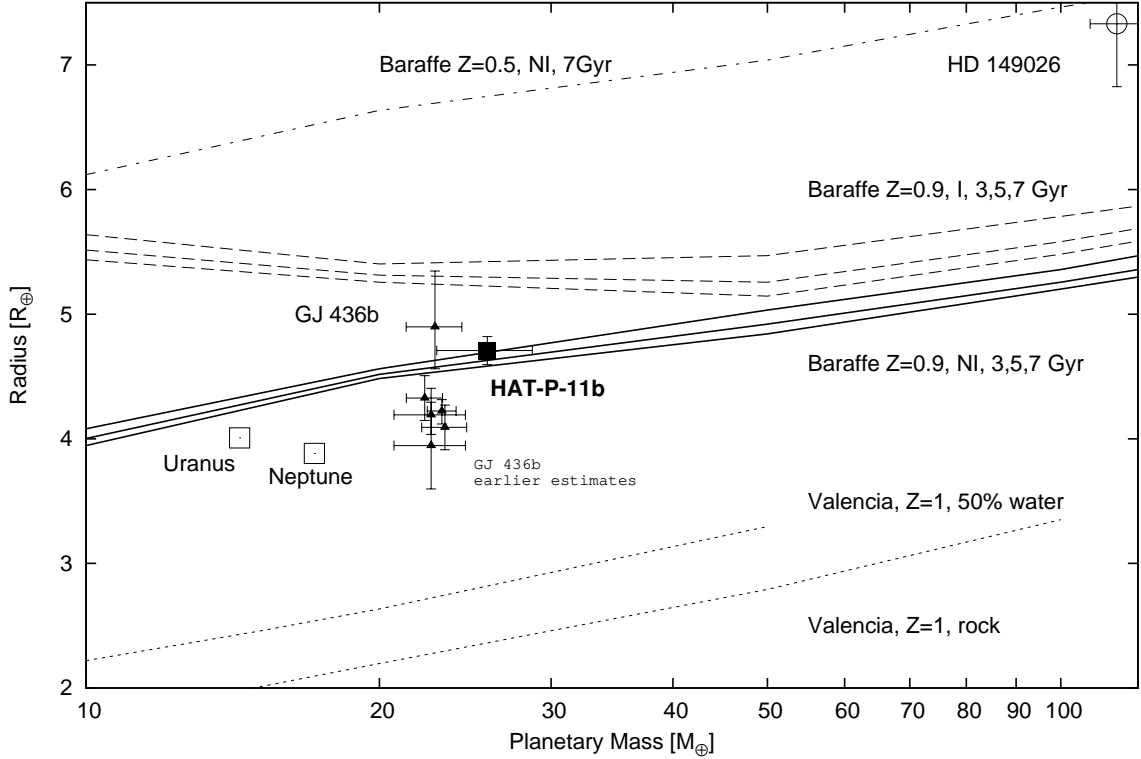}
\caption{
	The mass--radius diagram for Neptune-mass planets. Plotted are
	Uranus and Neptune from our own Solar system, the detection
	presented in this paper, \hatcurb, and the only other known
	transiting hot Neptune \gj{436b}. For \gj{436b} we plotted all
	available mass and radius determinations \citep{southworth:2008},
	the most recent one being the highest point in radius.  Also shown
	in the upper right corner is the hot Saturn \hd{149026b}. 
	Overplotted on the figure are theoretical models. \hatcurb\ is
	clearly much bigger than a $\sim \hatcurPPme\,\mearth$ rocky or
	water-rich ``super-earth'' would be, as based on models from
	\citet{valencia:2007}. \hatcurb\ is also inconsistent with
	irradiated extreme metal-rich ($Z=0.9$) models of
	\citet{baraffe:2008} (labeled as ``Baraffe Z=0.9, I, 3,5,7 Gyr'',
	``I'' stands for irradiated).  Indeed, these models assume
	$a_{equiv} = 0.045$\,AU semi-major axis around a solar twin,
	instead of the low irradiation received by \hatcurb ($a_{equiv} =
	0.1$\,AU). The $Z=0.5$ models from \citet{baraffe:2008} with a 50\%
	H/He envelope lie on the top of the figure, clearly presenting a
	mismatch with \hatcurb. However, \hatcurb\ is fully consistent with
	$Z=0.9$ metal-rich non-irradiated (``NI'') \citet{baraffe:2008}
	models (thick lines), in particular with the 3--5\,Gyr models.
\label{fig:nepmod}}
\end{figure*}

The stellar parameter determination (\refsec{stelparam}), the blend
analysis (\refsec{blend}) and the global modeling of the data presented
in \refsec{globmod} together show the detection of a
\hatcurPPme\,\mearth\ and \hatcurPPre\,\rearth\ planet that orbits a
bright ($V=\hatcurCCmag$) \hatcurYYspec\ dwarf star on an eccentric
($e=\hatcurRVeccen$) orbit with $P=\hatcurLCPshort$~day period, causing
a \hatcurRVK\,\ms\ RV variation of the host star and a
\hatcurLCdip\,mmag transit as it passes in front of the star.  The
planet is somewhat more massive than our own Uranus (14.3\,\mearth) and
Neptune (17.1\,\mearth), and slightly more massive than \gj{436b}
(22.6\,\mearth). Based on its mass and radius, it is justified to use
the metonym super-Neptune for classification of \hatcurb, just like for
\gj{436b}, the only similar object known so far. For the location of
$\sim$Neptune-mass objects, see \reffig{nepmod}.

When compared to models of \citet{fortney:2007}, \hatcurb\ is much
smaller in radius than similar mass planets with 50\% rock/ice core and
50\% H/He envelope, and \hatcurb\ is much bigger than pure rock/ice
planets without a gas envelope. The mismatch with such pure-rock/ice
``super-Earths'' is confirmed by models of \citet{valencia:2007}.
Overlaid in \reffig{nepmod} are super-Earth planet models with
Earth-like composition (lower dotted line) and 50\% by mass $\rm H_2 O$
content and Earth-like Si/Fe ratio (upper dotted line). Both models
fall clearly below the observational values for \hatcurb. As the
expected limit for gravitational capture of H/He is $\sim10\mearth$
\citep{rafikov:2006}, it is hardly surprising that \hatcurb\ has some
H/He, even if evaporation played an important role during its evolution.
\hatcurb\ receives only modest irradiation when compared to hot
Jupiters. The time-averaged flux over its eccentric orbit is
\hatcurPPfluxavg\,\ergscmsq, and the equivalent semi-major axis where
similar irradiation would be received from a Sun-like star is
$a_{equiv}\sim 0.1$\,AU. Using the models from \citet{fortney:2007},
the radius of \hatcurb\ is indeed consistent with a 90\% heavy element
planet with an irradiation corresponding to $a_{equiv} = 0.1$\,AU. In
these models, the heavy elements are located in the core, and the
envelope is metal-free.

We also compared the observational values to the thorough theoretical
work of \citet{baraffe:2008}. Planetary isochrones for various metal
content, age and irradiation are also plotted on \reffig{nepmod}. As
\citet{fortney:2007} note, planetary radii and their evolution are
scarcely affected when irradiation is small, and $a_{equiv} \gtrsim
0.1$\,AU, thus it is expected that \hatcurb\ will match non-irradiated
models better.  And indeed, the best match is with extreme metal rich
non-irradiated Baraffe models with $Z=0.9$ fraction of heavy elements
(rock and ice), and a 10\% H/He envelope. The non-irradiated metal-rich
models of various ages between 3 to 7\,Gyr lie very close to each
other, and are all within the observational error-bars. However, the
non-irradiated models with significant (50\%) H/He envelope (top of
figure) are far from matching the observations. \citet{baraffe:2008}
notes that in general, the distribution of the heavy elements (core
vs.~envelope) can have considerable impact on the planetary radius, but
in this extreme metal rich scenario, the freedom to vary their
distribution is limited.  Altogether, \hatcurb\ appears to be a
super-Neptune planet with $Z=0.9$ and with a 10\% H/He envelope.

It is interesting to compare \hatcurb\ to \gj{436b} --- the only other
super-Neptune with known mass and radius. These two planets are very
similar in some physical properties, in spite of the very different
environment. There are numerous system parameter determinations for
\gj{436b} available in the literature \citep[see compilation and Table
A11 of][]{southworth:2008}. Starting from the discovery of the transits
by \citet{gillon:2007} ($22.6\mearth$, $3.95\,\rearth$), planetary
masses range from $\mpl = 22.26\,\mearth$ to $24.8\,\mearth$, and
planetary radii range from $3.95\,\rearth$ to $4.9\,\rearth$
\citep{bean:2008a}. Some of these earlier estimates, including that of
\citet{bean:2008a} are plotted on \reffig{nepmod}. The RV amplitude of
\gj{436} ($K = 18.34\pm0.52$\,\ms, T08) is larger and better determined
than \hatcurb\ ($K=\hatcurRVK$\,\ms), and the transits are deeper
(6.5\,mmag vs.~4.2\,mmag), and have even been observed from space by
\ssts\ \citep{gillon:2007b,deming:2007} and \hst\ \citep{bean:2008a}.
The reason for the larger $K$ value and transit depth of \gj{436},
however, are due to the host star being a small mass and radius M-dwarf
\citep[$\mstar \approx 0.45\,\msun$, $\rstar \approx 0.46\,\rsun$, T08
and][]{southworth:2008}. Uncertainties in both the planetary mass and
radius remain considerable because of the uncertainty in the stellar
parameter determination for M dwarf stars (T08). Altogether, the masses
and radii of \gj{436b} and \hatcurb\ are very close (to within
2-$\sigma$). \gj{436b} has a slightly smaller and better determined
mass. \hatcurb\ has a comparable and somewhat better determined radius.

There are other similarities between the two planets. Both orbit on
short period mildly eccentric orbits, \gj{436b}: $e=0.16\pm0.02$
\citep{gillon:2007} and \hatcurb: $e = \hatcurRVeccen$. The fact that
they have not circularized yet is interesting, as the circularization
timescale appears shorter than the age of the host stars \citep[e.g.][
and references therein]{matsumura:2008}, although the stellar ages are
ill determined due to the unevolved M and K dwarf host stars.
%
As \citet{matsumura:2008} noted on the origins of eccentric close-in
planets, some of these planets may have larger than expected $Q$
specific dissipation functions (depending on the planetary structure,
and characteristics of the tides). The fact that neither of these hot
Neptunes has circularized yet may tell us something about the formation
and planetary structure of Neptunes. We note that the putative
HAT-P-11c planet (\refsec{secplan}) causing the long-term drift in the
data appears to have a large semi-major axis, because the RV drift does
not show any significant non-linear coefficient over the 1~year
timescale of the observations (\refsec{hires}). Thus, it is unlikely
that it is causing eccentricity pumping. Whether the Kozai mechanism
\citep{fabrycky:2007} plays an important role will be subject to
further investigations when the long-term drift in the RVs is better
characterized. Both \gj{436b} and \hatcurb\ have similar surface
gravities, $\loggpl = 3.11\pm0.04$ (cgs, T08) for \gj{436b}
vs.~$\loggpl = \hatcurPPlogg$ for \hatcurb. They also appear to have
similar mean densities, $\rhopl = 1.69\pm^{0.14}_{0.12}\,\gcmc$ (T08)
for \gj{436b} and $\hatcurPPrho\,\gcmc$ for \hatcurb.

It is surprising that in spite of these similarities (mass, radius,
density, surface gravity, eccentricity), the two planets are in grossly
different environments. Consequently, the \gj{436b}---\hatcurb\ pair
must yield tight constraints on theories that should reproduce
their properties in spite of the different environments. 
\gj{436b} orbits an M dwarf
with half the mass and radius of \hatcur, one tenth of the luminosity
($\lstar = 0.026\,\lsun$ vs.~$\hatcurYYlum$\,\lsun), and colder
effective temperature of $\teffstar = 3350 \pm 300$\,K (T08) versus
$\hatcurSMEteff$\,K for \hatcurb. The smaller semi-major axis of
\gj{436b} compensates to some extent for the lower luminosity host
star, but still, the irradiation they receive is grossly different
(\gj{436b}: $3.2\cdot10^{7}\ergscmsq$, \hatcurb:
$1.34\cdot10^{8}\,\ergscmsq$, i.e.~4 times more, as time-integrated
over the eccentric orbit). These lead to different expected equilibrium
temperatures: $650\pm60$\,K for \gj{436b} (T08) vs.~$\hatcurPPteff$\,K,
assuming complete heat redistribution. Also, the different stellar
effective temperatures lead to different spectral distributions of the
infalling flux.  It will be interesting to compare the atmospheric
properties of the two planets to see the effect of different integrated
flux and spectral flux distributions on these otherwise similar planets.

While both \gj{436b} and \hatcurb\ have similar mass and radius, and
inferred structure of $Z=0.9$ heavy element content with 10\% H/He
envelope, it is noteworthy that the metallicity of the host stars,
and presumably the environment these planets formed in, is different;
$\feh = -0.03 \pm 0.20$ for \gj{436} (T08) and $\feh = \hatcurSMEzfeh$
for \hatcur.

From the observational point of view, \hatcur\ is a bright star ($V\sim
9.58$), more than a magnitude brighter than \gj{436}. The transits of
\hatcurb\ are at a small impact parameter ($b=\hatcurLCimp$) as
compared to the near-grazing transit of \gj{436b} ($b = 0.848$). The
total duration of the transit for \hatcurb\ is about twice
(\hatcurLCdur~days) that of \gj{436b} (0.042\,d). We can thus expect
very high quality follow-up observations of \hatcurb\ in the future.

\subsection{A Second Planet in the System?}
\label{sec:secplan}

The radial velocity of \hatcur\ shows, in addition to the
\hatcurLCPshort~day period induced by \hatcurb\, a significant trend of
$G_1 = \hatcurRVlindrift\rm\,\ms\,day^{-1}$.
This drift could be due to
a second planet in the system, corresponding to 
$M_2\sin i_2/a_2^2 \sim \dot G_1 / G = (0.061 \pm 0.01)
\mjup\,\mathrm AU^{-2}$, where the subscript ``2'' refers to HAT-P-11c, and $G$
is the gravitational constant. 
Such a conjecture is scarcely
surprising: for example \cite{bouchy:2008} have noted that 16 out of
the 20 hot Neptune planets discovered heretofore are members of
multiple planet systems. Interestingly enough, as pointed out by
\cite{bouchy:2008}, this fraction (80\%) is significantly higher than
23\%, which is the fraction of all known exoplanets which are in
multiple systems. We also recall that the significant orbital
eccentricity of \hatcurb\ is reminiscent of that of \gj{436b}, and that
the latter has motivated the speculation that there may be another
planet in the system, sufficient to maintain the eccentricity of GJ436b
\citet{ribas:2008,bean:2008b}. Finally, as was also pointed out by
\cite{bouchy:2008}, of 7 multiple planet systems containing both a
gas-giant planet and a hot Neptune planet, all have super-solar
metallicities.  The high metallicity of \hatcur, $\feh =
\hatcurSMEzfeh$, then strengthens the suspicion that there may be
multiple planets in the system.

Continued monitoring of the radial velocity of \hatcur\ over the next
few years should indicate whether the drift persists, and 
whether it shows evidence of curvature of the orbit. Another way to
establish the presence of a second body in the system is to search for
transit timing variations in repeated transits
\citep{agol:2005,holman:2005}.

\subsection{Future Kepler Observations}

The fact that \hatcur\ will lie on one of the detector chips for the
Kepler spacecraft presents remarkable opportunities for scientific
followup of the discovery presented here.  If all goes well, Kepler
should be capable of 1-min cadence photometry at a precision of 0.1
mmag for this bright star. Currently Kepler is expected to be
operational for a minimum of 3.5 years, and perhaps longer; in that
time it should have more than 250 transits by \hatcurb. In addition it
should show photometric variability due to surface inhomogeneities such
as starspots, spanning more than 40 rotations if indeed the rotation
period of the planet is verified to be about 29.2 days. Here we list
some opportunities for important scientific followup with Kepler:

\paragraph{Kepler transit data}

With a depth of \hatcurLCdip\,mmag, the transit light curve of
\hatcurb\ should be measurable by Kepler with excellent precision for a
single light curve and extraordinary precision when many light curves
are combined.
%
%
To maximize the accuracy of parameters derived for  the \hatcur\ {\it
system}, it is important that, in addition to the Kepler observations,
precise ground-based radial velocity data be obtained through the
duration of the Kepler observations.

The possibility of a second planet in the system can be tested through
a search for transit timing variations (TTV's) in the large number of
transits to be observed by Kepler. As noted above, radial velocity data
obtained over the same time frame could provide independent evidence
for such a planet, and comparison of the two data sets would be of
obvious importance.

\paragraph{Kepler Out-of-transit photometry} 

The Kepler data should reveal the detailed modulation of photospheric
brightness--presumably due to starspot activity--with great precision
and time coverage over the lifetime of the mission. This could yield
much useful information about the star, including its rotation and even
differential rotation \citep{frohlich:2007}. Furthermore, if multiple
transits of \hatcurb\ pass in front of the same star-spot on the face
of the slowly rotating star, then extraordinary precision can be
achieved in determining the rotation rate \citep{sv:2008}. Comparison
with concurrent ground-based measurements of the chromospheric emission
$S$ index will enable investigation of how photospheric spots correlate
with chromospheric emission on stars other than the sun.  As noted
earlier, the data on $S$ index reported here show a long-term ($\sim$
450 day) variation reminiscent of solar activity cycles; monitoring
both photospheric emission variations from Kepler broadband photometry
and chromospheric emission variations from ground based telescopes
could prove useful.

\paragraph{Kepler asteroseismology}

Kepler will obtain extraordinary information about the internal
structure and evolution of stars from asteroseismic investigations,
which measure the frequencies and amplitudes of acoustic oscillation
modes (``p-modes'') through small changes in the integrated brightness
of the star. The ``large splitting'' of p-mode frequencies yields
direct information on the radius of the star, and their ``small
splitting'' gives information on the H/He ratio in the core and hence
the age of the star \citep[e.g.][]{cd:2004}.  A great deal of
additional information is possible if the star is accompanied by a
transiting planet (like \hatcurb): in this case the transit light curves
yield completely independent information on the radius and mean density
of the star. An additional bonus in the case of \hatcur\ is that it has
an excellent Hipparcos parallax. Combining these independent pieces of
information should lead to very detailed knowledge of its internal
structure, including second-order near-surface effects not accounted
for in normal interior structure models (Kjeldsen et al 2008).
Information on the age of the star, if it can be obtained from
measurement of the small splitting, could provide a useful comparison
with the age inferred either from the global (\lc, RV, stellar
isochrone) modeling of the data or from the Ca HK-index.

As noted above, there is a suggestion from the variation of its $S$ index
that \hatcur\ is undergoing long-term variations of its surface
magnetic activity somewhat reminiscent of the solar activity cycle. On
the Sun, solar cycle-related variations of solar magnetic activity are
known to be correlated with frequency variations of p-modes
\citep[e.g.][]{woodard:1985,woodard:1991}, because the sub-surface
structure is modified in regions of solar magnetic activity. 
Monitoring the p-mode frequencies of \hatcur\ over the course of the
Kepler mission, along with continued ground-based monitoring of the
S-index, may show whether similar behavior occurs in the sub-surface
layers of another star with significantly different structure.

Finally, we note that two other stars with known transiting planets,
namely TrES-2 \citep{odonovan:2006} and HAT-P-7 \citep{pal:2008a}, also
lie on the {\it Kepler} detector array, so it should be possible to
carry out studies of these stars and their planets similar to those
described above for the \hatcur\ system. It would be interesting to
compare the results for these two stars, with spectral type F and G
respectively, with those for the K star \hatcur.  Almost certainly many
more TEP systems will be discovered by Kepler, but those systems that
are known in advance of the mission should be especially valuable since
observations with the 1-minute cadence can be scheduled from the start
of the mission, plus the longer baseline improves our sensitivity to
detect secular changes. 


\acknowledgements 

HATNet operations have been funded by NASA grants NEG04GN74G,
NNX08AF23G and SAO IR\&D grants. Work by G.\'A.B.~and J.~Johnson was
supported by Postdoctoral Fellowships of the NSF Astronomy and
Astrophysics Program (AST-0702843 and AST-0702821, respectively). We
acknowledge partial support also from the Kepler Mission under NASA
Cooperative Agreement NCC2-1390 (D.W.L., PI). G.K.~thanks the Hungarian
Scientific Research Foundation (OTKA) for support through grant
K-60750. This research has made use of Keck telescope time granted
through NOAO (program A285Hr) and NASA (N128Hr).

Data presented herein were obtained at the W.~M.~Keck Observatory from
telescope time allocated to NASA through the agency’s scientific
partnership with the California Institute of Technology and the
University of California. The Observatory was made possible by the
generous financial support of the W. M. Keck Foundation.”





\appendix

\section{Systematic variations}
\label{sec:sysvar}

As regards the red-noise or systematic variations (also referred to as
``trends''), we may distinguish between those effects that have a well
understood reason, and where deviations correlate with a set of
external parameters, and other systematic variations where the
underlying parameters are not known. The deviations are usually taken
with respect to the median of the \lc.

The $m(t_i)$ \lc\ 
of a star observed at $t_i$ time instances can be decomposed as:
\begin{equation}
	m(t_i) = m_0(\vec{p},t_i) + G(t_i) + E(\vec{e_i}(t_i)) + T(t_i),
	\label{eq:lc}
\end{equation}
where $m_0(\vec{p},t_i)$ is the systematics-free \lc\ as a function of
time, and parametrized by a discrete set of parameters denoted as
$\vec{p}$, $G(t_i)$ is white noise, $E$ is the systematic variation due
to the change in external parameters $\vec{e_i}$, and $T$ denotes the
additional trend that is {\em not} known to be simple function of such
parameters. For planetary transits the functional form of $m_0$ can be a
transit model by \cite{mandel:2002}, and the $\vec{p}$ parameters can
be the $\arstar$ geometric semi-major axis, $p\equiv\rstar/\rpl$
relative diameter of the planet, the normalized impact parameter $b$
and the $T_c$ center of transit. For a \dscu\ variable star, these
parameters can be the amplitudes and phases of various Fourier components.

\subsection{External Parameter Decorrelation}
\label{sec:epd}

Although briefly described earlier \citep{bakos:2007b}, it is worth
defining the External Parameter Decorrelation (EPD) technique, since it
is extensively used in this work.  The EPD method attempts to determine
the actual functional form of $E$ in \refeq{lc}. The EPD effects are
treated and determined as specific for each star, i.e.~no information
from other stars are used (this is a key difference when compared to
TFA that uses a template of other stars --- see later).

\paragraph{Constant (or simple) EPD:} The simplest form of EPD is when we assume
that the underlying $m_0$ signal is constant, and also that $\langle E
\rangle \gtrsim \langle T \rangle$ (i.e.~the effects to be corrected by
EPD are of the same order as those corrected by TFA). A typical
application is for planetary transits in survey mode, since such
transits are short events compared to the total orbital time, and have
mostly been observed around stars with much smaller variation than the
transit amplitude \citep[note that there are exceptions;
see][]{alonso:2008}, thus the underlying signal can be approximated
with its median $\sim$95\% of the time. The $E$ relation between
$\delta m (t_i) \equiv 
	m(t_i) - m_0(t_i) = 
	m(t_i) - \langle m(t_i) \rangle_m$
and $\vec{e_i(t_i)}$ parameters are sought via a chi-square
minimization, usually by non-linear least squares method with outlier
rejection ($\langle \rangle_m$ denotes median value). Upon determining
$E$, we get an EPD corrected signal 
$\delta m_{EPD}(t_i) = \delta m (t_i) - E(t_i)$. 

Our experience with HATNet survey data has shown that the effects of
$E$ and $T$ are indeed comparable, and thus we use constant EPD.
Specifically, the external parameters are the X,Y sub-pixel position,
the background and its standard deviation, the stellar profile
parameters characterizing the PSF width (called $S$) and its elongation
($D$ and $K$), plus the hour-angle and zenith distance.

\paragraph{Simultaneous EPD:} Used primarily in the analysis of the
photometric follow-up data, where most of the observations are centered
on the transits, thus the assumption of constant signal does not hold.
One possibility is to use the out-of-transit (OOT) section of the \lc\
to determine the EPD parameters in constant EPD mode, and then apply
the correction for the in-transit section of the \lc\ as well. This is
sub-optimal, as the OOT may be very short compared to the in-transit
section, or may be missing.  Thus, the key difference compared to
constant EPD is that we use all data to recover the dependence on
external parameters, i.e.~$m_0(\vec{p},t_i)$ in \refeq{lc} is not
constant, but assumed to be a function of other parameters and time.
The fitting procedure determines both the $E$ correlation and the
optimal set of $\vec{p}$ parameters simultaneously.

\subsection{The Trend Filtering Algorithm}
\label{sec:tfa}

The signal after (or without) the EPD procedure still contains the
general systematic variations denoted as $T(t_i)$ in \refeq{lc}.
These are suppressed by the Trend Filtering Algorithm \citep[TFA,
see][]{kovacs:2005}, assuming that certain other stars in an $M$ element
\lc\ ``template'' show similar variations.

To recall, for selected star $j$, TFA minimizes the following expression:
\begin{equation}
	\label{eq:tfamin}
	D = \sum_{i=1}^{N}
	\big\lgroup
			m_j(t_i) - m_{j,0}(\vec{p},t_i) - F_M(t_i) -
			E_j(\vec{e}(t_i))
	\big\rgroup^2,
\end{equation}
where
\begin{equation}
	\label{eq:tfafilter}
	F_M(t_i) = \sum_{k=1,k\neq j}^{k=M}{c_k m_k(t_i)}.
\end{equation}
is the TFA filter function of M elements. The notation is like above in
\refeq{lc}, $m_{j,0}(\vec{p},t_i)$ is the model function and
$E_j$ is the EPD correction. 

There are many variants of TFA, and they have been used widely in this
work. The signal search in the HATNet data is done in two parallel
steps. First via {\em simple (or constant) TFA} where there is no
assumption on the periodicity, the model function $m_{j,0}$ is a
constant, $E_j$ is zero, as EPD has been performed as an independent
step before TFA, and the $c_k$ coefficients are sought.
Second, via {\em reconstructive TFA}, where the simple TFA is followed
by a frequency search, the signal is phase-folded with the most
significant frequency, the model function $m_{j,0}$ is fitted to the
folded data, the model function is un-wrapped to the original
time-base, and $D$ is minimized again to iteratively determine the
$c_k$ coefficients. 
The third method, introduced in this work, is {\em simultaneous TFA},
whereby the $c_k$ TFA coefficients and the functional dependence of
$m_{j}(\vec{p},t_i)$ on $\vec{p}$ parameters are simultaneously
determined. 

EPD and TFA can be performed sequentially, or even simultaneously. In
the analysis of the photometric follow-up data (see \refsec{phfu}) we
implemented such a {\em simultaneous EPD-TFA}, where the $E_j(\vec{e})$
EPD function is simultaneously fitted with the $c_k$ TFA template
coefficients and the $m_{j}$ model function.

Both TFA and EPD can be performed {\em globally}, using one set of
coefficients for the entire dataset, and {\em locally}, when
\refeq{tfamin} is split up into smaller data-blocks, such as one-night
segments, and the $c_k$ TFA coefficients and the EPD function
parameters are fitted for each data-block separately to allow for
changing systematics. Conversely, for {\em global TFA} systematic
variations of template stars have to ``match'' those of the main target
(minus the model) for all nights.

Finally, we note that if we do not use EPD (eliminate the term for $E$
in \refeq{lc}), but apply TFA only, then naturally, TFA will take
care of some of the systematics that otherwise would have been
corrected by EPD.

\end{document}